\documentclass[british,a4paper,10pt]{article}
\usepackage{amssymb}
\usepackage{amsmath}
\usepackage{subfigure}
\usepackage{slashed}
\usepackage{amsmath}
\usepackage[normalem]{ulem}
\usepackage{graphicx,psfrag}
\usepackage{mathrsfs}
\usepackage{amsfonts}
\usepackage{multirow}
\usepackage{hyperref}
\usepackage{placeins}
\usepackage{amsthm}
\usepackage{latexsym}
\usepackage[dvips]{epsfig}
\usepackage{enumitem}  
\usepackage{soul}
\usepackage{graphicx}
\usepackage{amsthm}
\usepackage{latexsym}
\usepackage[dvips]{epsfig}
\usepackage[utf8]{inputenc}
\usepackage{babel}
\usepackage{multicol}
\usepackage{color}
\usepackage{comment}
\usepackage[dvipsnames]{xcolor}
\usepackage{wasysym}
\usepackage{mathrsfs}
\usepackage{bm}
\usepackage{authblk}
\usepackage{slashed}
\usepackage{yhmath} 
\usepackage{stmaryrd}

\usepackage{pifont}
\usepackage{mdframed}
\theoremstyle{plain}
\newtheorem{proposition}{Proposition}

\newtheorem{assumption}{Assumption}

\newtheorem{remark}{Remark}
\newenvironment{IndRemark}{
  \begin{center}
  \begin{minipage}{0.95\textwidth}
  \begin{remark}
}
{
\end{remark}
\end{minipage}
\end{center}
}
\newenvironment{IndAssumption}{
  \begin{center}
  \begin{minipage}{0.95\textwidth}
  \begin{assumption}
}
{
\end{assumption}
\end{minipage}
\end{center}
}
\newenvironment{IndProposition}{
  \begin{center}
  \begin{minipage}{0.95\textwidth}
  \begin{proposition}
}
{
\end{proposition}
\end{minipage}
\end{center}
}
\newcommand\reducedbox{\vcenter{\hbox{\scalebox{0.5}{$\fbox{\ding{52}}$}}}}

\def\bmeta{{\bm \eta}}
\def\bmg{{\bm g}}

\def\bmp{{\bm p}}
\def\bmf{{\bm f}}
\def\bmF{{\bm F}}
\def\bmT{{\bm T}}
\def\bmQ{{\bm Q}}
\def\bmS{{\bm S}}

\def\bmH{{\bm H}}

\def\dotnabla{{\mathring{\nabla}}}
\def\dotg{{\mathring{g}}}

\def\dotXi{{\mathring{\Xi}}}
\def\dotd{{\mathring{d}}}
\def\dotPhi{{\mathring{\Phi}}}
\def\dotT{{\mathring{T}}}
\def\dotF{{\mathring{F}}}

\usepackage[symbol]{footmisc}

\allowdisplaybreaks
\begin{document}

\title{\textbf{Linearised conformal Einstein field equations}}
\author[1]{Justin Feng  \footnote{E-mail address:{\tt justin.feng@tecnico.ulisboa.pt}}}
\author[2]{Edgar Gasper\'in *   
  \footnote{E-mail address:{\tt edgar.gasperin@tecnico.ulisboa.pt}}}
  \affil[1,2] { CENTRA, Departamento de F\'isica, Instituto Superior
    T\'ecnico IST, Universidade de Lisboa UL, Avenida Rovisco Pais 1,
    1049 Lisboa, Portugal \\ * edgar.gasperin@tecnico.ulisboa.pt }

  \maketitle

  \vspace{-4mm}
  
\begin{abstract}
  The linearisation of a second-order formulation of the conformal
  Einstein field equations (CEFEs) in Generalised Harmonic Gauge (GHG),
  with trace-free matter is derived.  The linearised equations are
  obtained for a general background and then particularised for the
  study linear perturbations around a flat background ---the inversion
  (conformal) representation of the Minkowski spacetime--- and the
  solutions discussed. We show that the generalised Lorenz gauge
  (defined as the linear analogue of the GHG-gauge) propagates.
  Moreover, the equation for the conformal factor can be trivialised
  with an appropriate choice for the gauge source functions; this
  permits a scri-fixing strategy using gauge source functions for the
  linearised wave-like CEFE-GHG, which can in principle be generalised
  to the nonlinear case. As a particular application of the linearised
  equations, the far-field and compact source approximation is employed
  to derive quadrupole-like formulae for various conformal fields such
  as the perturbation of the rescaled Weyl tensor.
\end{abstract}


\section{Introduction}

Although nowadays gravitational wave effects can be studied using the
full non-linear Einstein field equations, linear perturbation theory
remains an active area of research, e.g. black hole perturbation
theory, quasinormal modes, post-Newtonian and post-Minkowskian
expansions.  As it is well-known, the concept of gravitational radiation
can be traced back to 1918 to the study of the linearisation of the
Einstein field equations around flat spacetime and the derivation of the
celebrated quadrupole formula. However, a rigorous formulation of the
notion of gravitational radiation had to wait for further development of
mathematical aspects of the theory put forward by Bondi, Sachs, Newman
and Penrose and others in the decade of 1960 ---see \cite{HilNur16,
BieGarYun17} for a historical review.  The notion of null-infinity (and
more generally the notion of a conformal boundary) developed in the
aforementioned theory proved crucial for the modern understanding of
gravitational radiation.

\medskip

Null-infinity $\mathscr{I}$ corresponds to an idealisation of the
asymptotic region of the physical spacetime $(\tilde{\mathcal{M}},
\tilde{\bmg})$ and it is defined through a conformally related manifold
$(\mathcal{M},\bmg)$ where $\bmg=\Xi^2\tilde{\bmg}$ so that
$\mathscr{I}$ can be identified with the region where $\Xi=0$ ---but
$\nabla_a \Xi \neq 0$.  Unfortunately, the Einstein field equations
(EFEs) are not conformally invariant so evaluating quantities at
$\mathscr{I}$ is non-trivial. To circumvent this problem, there are
different approaches aimed at specific goals. For instance, the
hyperboloidal approach in their different renditions \cite{Zen08,
VanHusHil14, Van15, VanHus17, HilHarBug16, GasHil18, DuaFenGasHil22}
represents a compromise between the preservation of standard
formulations of the EFEs used in Numerical Relativity and the inclusion
of  $\mathscr{I}$ at the expense of dealing with formally singular (but
numerically tractable) equations.  On the other hand, there exists
another strand of research that stems from taking further the conformal
approach while insisting on having formally regular equations.  This
conformally regular approach was initiated in 1981 when H. Friedrich
derived in \cite{Fri81} a reformulation of the Einstein field equations
known as the conformal Einstein field equations (CEFEs) ---see
\cite{Val16, Fra04} for an exhaustive discussion. One of the distincive
features of the CEFEs is that the conformal factor is a variable, hence
the location of the conformal boundary is not fixed a priori. Strategies
for fixing the conformal boundary, or \emph{scri-fixing}, have been
introduced for the CEFEs in \cite{Fri95} for vacuum and \cite{LueVal12}
for trace-free matter, but they require the existence of a congruence of
conformal curves.

\medskip

Null-infinity is one of the central concepts that go into some of the
most important open problems in General Relativity such as the weak
cosmic censorship conjecture and the global stability analysis of
spacetimes.  Additionally, it is crucial for gravitational waves: due to
the non-localisability of gravitational radiation, it is only rigorously
defined at null-infinity.  The CEFEs are a reformulation of the Einstein
field equations that incorporates Penrose's conformal compactification
into the initial value problem in General Relativity. Hence, the
systematic study of these equations is important for the deeper
understanding of gravitational radiation and related effects such as the
memory effect. However, despite many advances in the mathematical
analysis of spacetimes and numerical evolutions with the CEFEs
\cite{Fri02, LueVal09, LueVal12, GasVal17, DouFra16, Hub99, Hub01,
Hus02b}, the application of the CEFEs to physical problems has been
surprisingly limited. This is evident in (and possibly a result of) the
fact that the application of linear perturbation theory to the CEFEs has
not been studied in the existing literature. Such an analysis is needed
to make contact with standard (linear) metric approaches to
characterising gravitational radiation. This article aims to provide a
first step in this direction by studying the linearisation of the CEFEs
in a form that resembles standard formulations of the linearised
Einstein field equations. This article represents step in exploiting the
CEFEs for physical applications. In the current article, we focus on
linearisations around flat spacetime however in the future we plan to
extend our analysis into more interesting backgrounds and derive the
conformal (meaning derived from the Conformal Einstein field equations)
counterpart of black hole perturbation theory and post-Newtonian
expansions.

\medskip

The core of the CEFEs are the Bianchi identities, which provide a set of
evolution and constraint equations for the Weyl curvature (coupled with
other fields in the non-linear case). One approach to the linearised
problem concerns the study of the spin-2 equation in a fixed background
spacetime, which can be thought as the linearisation of the Weyl sector
of the CEFEs in spinorial form. The spin-2 equation on a particular
background known as the Minkowski $i^0$-cylinder has been analysed in
\cite{Val03a, BeyDouFra13, Fri02}.  The spin-2 equation
$\nabla_{A'}{}^A\phi_{ABCD}=0$, although elegant, looks very different
to the standard linearisation of metric formulations of the Einstein
field equations.  Hence, in this article, we study the linearisation not
of the original set of CEFEs but rather we start from a metric and
second-order formulation of the CEFEs which is closer in spirit to
standard hyperbolic reductions of the EFEs in generalised harmonic gauge
(GHG) employed in Numerical Relativity. This non-linear wave-like
formulation of the CEFEs was originally derived in the vacuum case in
\cite{Pae13} and has been extended for the case of trace-free matter in
\cite{CarHurVal}. In this paper, we linearise these equations around a
general background and then the solutions to the linearised equations
around flat spacetime are studied. As a concrete application of the
linearisation, we derive, in the far-field approximation,
quadrupole-like formulae for the conformal fields, obtaining in
particular a quadrupole-like formula for the perturbation of the
rescaled Weyl tensor. As a byproduct, we show how the gauge source
functions can be chosen so that the conformal factor remains
unperturbed. This can be regarded as a ``scri-fixing'' strategy for the
linearised CEFEs for trace-free matter. We then argue that the same
strategy carries over to the non-linear wave-CEFEs in the context of the
hyperboloidal initial value problem, providing a simpler alternative to
the scri-fixing strategy in \cite{LueVal12} based on conformal
curves.

\subsection*{Notations and conventions}

The signature convention for (Lorentzian) spacetime metrics will be $
(-,+,+,+)$. Latin indices from the first half of the alphabet
$a,b,c,,...$ will be used as abstract tensor indices while Greek indices
will denote spacetime coordinate indices, taking values from the set
$\{0,1,2,3\}$.  If an adapted coordinate system (with coordinates
$x^\mu$) is introduced, $x^0=t$ represents the time coordinate, and
Latin indices from the second half of the alphabet represent spatial
indices, taking values from the set $\{1,2,3\}$. Symmetrisation and
antisymmetrisation of tensors will be denoted with round and square
brackets and taking the trace-free part of a tensor will be denoted by
adding $\{\}$ on the relevant indices. For instance $T_{(ab)}:=
\tfrac{1}{2}(T_{ab} + T_{ba})$, $T_{[ab]}:=\tfrac{1}{2}(T_{ab} -
T_{ba})$ and $T_{\{ab\}}:= T_{ab} - \tfrac{1}{4}g_{ab}T$.  The curvature
conventions are fixed by the relation
\[
(\nabla_a \nabla_b -\nabla_b \nabla_a) v^c = R^c{}_{dab} v^d.
\]
Although we will occasionally refer to expressions in spinorial form (as
much of the literature on the CEFEs employs spinor notation), all
calculations in this article are performed in tensor notation.

\subsection*{Organisation of the paper}

Section \ref{Sec:CFEs} gives a brief summary of the CEFEs and the
wave-like hyperbolic reduction of \cite{CarHurVal}. This is included to
give a concise reference to the reader and to provide the non-linear
context of the equations studied in this paper.  Section
\ref{sec:lin_wave_CEFE_gen} contains the linearisation of the wave-CEFEs
in GHG around a general background.  Section \ref{sec:Wave_CEFE_flat}
particularises the equations for a flat background and discusses the
perturbed solutions via Green's functions methods and, in the far-field
approximation, quadrupole-like formulae for the conformal fields are
derived.

\section{The conformal Einstein field equations}
\label{Sec:CFEs}

The conformal Einstein field equations (CEFEs) are a reformulation of
the Einstein field equations (EFEs) with the aim of dynamically
implementing R. Penrose's conformal approach. The CEFEs were originally
introduced in \cite{Fri81a} and encode a set of differential conditions
satisfied by the geometry of a conformal extension $(\mathcal{M},\bmg)$
of a spacetime $(\tilde{\mathcal{M}},\tilde{\bmg})$ (with
$\bmg=\Xi^2\tilde{\bmg}$) satisfying the EFEs. To differentiate between
these two, the pair $(\mathcal{M},\bmg)$ is called the \emph{unphysical
spacetime} while the pair $(\tilde{\mathcal{M}},\tilde{\bmg})$ is called
the \emph{physical spacetime}.  The feature that distinguishes the CEFEs
from other reformulations of the EFEs is that (for trace-free matter)
the CEFEs are \emph{formally regular} at $\Xi=0$ ---see
\cite{Val16,Fri81a,Fri81b}. To see this and concisely introduce the
CEFEs, it is convenient to define the following \emph{zero-quantities}
---see \cite{Val16} for an extensive discussion:
\begin{subequations}\label{CFE_tensor_zeroquants}
\begin{eqnarray}
  && Z_{ab} := \nabla_{a}\nabla_{b}\Xi +\Xi L_{ab} - s g_{ab} -
  \frac{1}{2}\Xi^3T_{\{ab\}}=0 ,
  \label{StandardCEFEsecondderivativeCF}\\
 && Z_{a} := \nabla_{a}s +L_{ac} \nabla ^{c}\Xi -
 \frac{1}{2}\Xi^2T_{\{ab\}}\nabla^b\Xi -
 \frac{1}{6}\Xi^3\nabla^{b}T_{\{ba\}}=0 , \label{standardCEFEs}\\ &&
 \delta_{bac} := \nabla_{a}L_{bc}-\nabla_{b}L_{ac} -
 d^{d}{}_{cab}\nabla_d{}\Xi -\Xi T_{abc} =0 ,
 \label{standardCEFESchouten}\\ && \lambda_{abc}:=
 \nabla_{e}d^{e}{}_{abc}-T_{bca}=0 , \label{standardCEFErescaledWeyl}\\
 && P^c{}_{dab} :=  R^c{}_{dab} - \Xi d^c{}_{dab} - 
 2( \delta^c{}_{[a}L_{b]d} -
   g_{d[a}L_{b]}{}^c).
 \label{standardCEFERiemannDecomp}\\
 && Z := \lambda - 6 \Xi s + 3 \nabla_{a}\Xi \nabla^{a}\Xi
 -\frac{1}{4}\Xi^4T=0\\
 \label{standardCFEconstraintFriedrichScalar}
  && M_a:=\nabla^bT_{ba} - \Xi^{-1}T\nabla_a\Xi=0.
\end{eqnarray}
\end{subequations}
where $\Xi$ is the conformal factor, $s$ is the \emph{Friedrich scalar},
$L_{ab}$ is the \emph{Schouten tensor}, $R^a{}_{bcd}$ is the Riemann
tensor, $d^{a}{}_{bcd}$ is the \emph{rescaled Weyl tensor}, $T_{ab}$ is
the \emph{rescaled energy-momentum tensor} with trace $T:=g^{ab}T_{ab}$
and $T_{abc}$ is the \emph{rescaled Cotton tensor}. The zero-quantities
$Z_{ab}, Z_a, \delta_{abc},\lambda_{abc}$ , $Z$ and $M_a$ are defined
simply as a bookkeeping device in the sense that the CEFEs are satisfied
when a collection of fields
\[
\{g_{ab}, \; \Xi, \; s\;,L_{ab},\; d_{abcd},\; T_{ab}, \; T_{abc}\}
\]
solve the equations
\begin{equation}\label{vanishing_CFEs_tensorial_zq}
  Z_{ab}=0, \quad Z_{a}=0, \quad \delta_{abc}=0, \quad
  \lambda_{abc}=0,  \quad P^{d}{}_{abc}=0, \quad Z=0, \quad M_a=0.
\end{equation}
\begin{IndRemark}
  \emph{ Observe that the only singular terms in the zero-quantities
  \eqref{CFE_tensor_zeroquants} appear with the trace of the unphysical
  energy-momentum tensor, specifically in equation
  \eqref{standardCFEconstraintFriedrichScalar}. However for trace-free
  matter all the equations implied by the zero-quantities
  \eqref{CFE_tensor_zeroquants} are formally regular at $\Xi=0$. }
\end{IndRemark}

\noindent The geometric variables are defined via
\begin{align}\label{defs_conf_vars}
  s&:= \tfrac{1}{4}\nabla_{a}\nabla^{a}\Xi +
  \tfrac{1}{24}R\Xi,\\ L_{ab}&:=\tfrac{1}{2}R_{ab}-\tfrac{1}{12}Rg_{ab},
  \\ d^{a}{}_{bcd}&:=\Xi^{-1}C^{a}{}_{bcd}.
\end{align}
where $R_{ab}$ and $R$ denote the Ricci tensor and Ricci scalar of
$(\mathcal{M},\bmg)$ while $C^{a}{}_{bcd}$ is the conformally invariant
Weyl tensor. Given a solution to the CEFEs, it can be verified that the

associated physical metric $\tilde{g}_{ab}=\Xi^{-2}g_{ab}$ will satisfy
the Einstein field equations:
\begin{align}
  \tilde{R}_{ab}-\frac{1}{2}\tilde{R}\tilde{g}_{ab} + \lambda
  \tilde{g}_{ab}= \tilde{T}_{ab} ,
\end{align} 
where $\tilde{T}_{ab}$ is the physical energy-momentum tensor, $\lambda$
is the cosmological constant, and $\tilde{R}_{ab}$ and $\tilde{R}$ are
the respective Ricci tensor and Ricci scalar. The matter variables in
\eqref{CFE_tensor_zeroquants} are related to their physical counterparts
as:
\begin{align} 
  T_{ab}& :=\Xi^{-2}\tilde{T}_{ab}, \qquad T:=\Xi^{-4}\tilde{T},
  \qquad \tilde{T}:=\tilde{g}^{ab}\tilde{T}_{ab} ,
  \\
  T_{abc}& :=\Xi^{-1}\tilde{\nabla}_{[a}\tilde{T}_{b]c}
  -\frac{1}{3}\Xi^{-1}\tilde{g}_{c[b}\tilde{\nabla}_{a]}\tilde{T} .
\end{align}
Here $\tilde{T}_{ab}$ is the physical energy-momentum tensor and
$\tilde{\nabla}$ is the Levi-Civita connection of $\tilde{g}_{ab}$. A
comprehensive discussion and derivation of the CEFEs can be found in
\cite{Val16}.

\begin{IndRemark}
  \emph{ Though the conceptually clean setup is the vaccum case, we
  include matter primarily for the purpose of deriving quadrupole-like
  formulae in subsection \ref{sec:quadrupolelike}.}
\end{IndRemark}

\begin{IndAssumption}
  \emph{
  From this point onward, it will be assumed that only trace-free
  matter models are in consideration. Notice that the conditions $T=0$
  and $T_{\{ab\}}=T_{ab}$ simplify some of the matter terms in the
  zero-quantities \eqref{CFE_tensor_zeroquants}.
  }
\end{IndAssumption}

We now consider some properties of the rescaled Cotton tensor under the
trace-free matter assumption. Under this assumption, the rescaled Cotton
tensor can be written in terms of the unphysical energy-momentum tensor
as
\begin{align}
  T_{abc}& :=\Xi \nabla_{[a}T_{b]c}
  + 3 T_{c[b}\nabla_{a]}\Xi -g_{c[a}T_{b]e}\nabla^e\Xi .
  \label{eq:CottonExpression}
\end{align}
Observe as well that the Cotton tensor has the symmetries
$T_{abc}=T_{[ab]c}$ and $ T_{[abc]}=0$. Furthermore, the trace-free
assumption and $\tilde{\nabla}^a\tilde{T}_{ab}=0$ implies that
\begin{align}\label{eq:unphysicalEMconservation}
  \nabla^aT_{ab}=0, \qquad
  \nabla_cT_{ab}{}^c & = 0. \nonumber
\end{align}
Additionally, using again equations \eqref{eq:CottonExpression},
\eqref {eq:unphysicalEMconservation}
 and  exploiting $Z_{ab}=0$ gives
\begin{align}
  \nabla _{a}T_{b}{}^{a}{}_{c} =&
  - \tfrac{1}{2} \Xi  \nabla _{a}\nabla ^{a}T_{bc}
  - (4 s - \tfrac{1}{3} \Xi  R)T_{bc}  +  \Xi ^3 T_{b}{}^{a} T_{ca}
  - \tfrac{1}{2} \Xi ^2 T^{ad} d_{bacd}  
   - \tfrac{1}{4} \Xi ^3 T_{ad} T^{ad} g_{bc} \nonumber \\ &
  + \nabla ^{a}\Xi  (-2 \nabla _{a}T_{bc} +  \nabla _{(b}T_{c)a}).
\end{align}

\begin{IndRemark}\label{remarkConfGauge}\emph{
  The Ricci scalar $R$ is not determined by the CEFEs and in fact, it
  encodes the conformal gauge freedom. Given two conformal extensions of
  the same physical spacetime  $\bmg=\Xi^2\tilde{\bmg}$ and
  $\check\bmg=\check{\Xi}^2\tilde{\bmg}$ one has that
  $\check{\bmg}=\varkappa^2 \bmg$ with $\varkappa := \check{\Xi}/\Xi$
  and $\varkappa \simeq \mathcal{O}(1)$ at the conformal boundary.
  Hence, their Ricci scalars are related via
  \begin{align}\label{conformalGaugeFreedom}
  6 \nabla_a \nabla^a \varkappa -R \varkappa =- \check{R}\varkappa^3.
  \end{align}
  Observe that if $\check{R}$ is considered as a given scalar function in
  $(\mathcal{M},\bmg)$ then, this equation can always be solved locally
  for $\varkappa$. Thus, the Ricci scalar in the CEFEs is a gauge
  quantity that is called the \emph{conformal gauge source function}
  ---see \cite{Val16} for further discussion.
  }
\end{IndRemark}

In this article, we will adhere to the metric formulation of the CEFEs
which consists of using the definition of the Schouten tensor to obtain
 an equation for the metric. Namely, from
\begin{align}\label{Ricci_ToSchouten}
R_{ab}=2L_{ab} + \frac{1}{6}Rg_{ab},
\end{align}
one substitutes the expression for the Ricci tensor as second
derivatives of the metric respect to some coordinate basis $x^\mu$ and
reads $L_{\mu\nu}$ as a source term in the resulting equation. The
source term $L_{\mu\nu}$ is then coupled to the rest of the variables
via equations \eqref{vanishing_CFEs_tensorial_zq}.

\begin{IndRemark}
  \emph{
   In view that the Ricci scalar is the conformal gauge source function,
   to have a clean split between gauge quantities and non-gauge
   quantities one could opt for using the trace-free Ricci tensor as
   variable instead of $L_{ab}$,
  \begin{align}
    \Phi_{ab} := \frac{1}{2}\Big(R_{ab} - \frac{1}{4}Rg_{ab}\Big).
  \end{align}
  The factor $1/2$ in its definition is conventional and it is put so
  that $\Phi_{ab}$ corresponds to the tensorial counterpart of the
  trace-free Ricci spinor $\Phi_{AA'BB'}$ of the Newman-Penrose
  formalism. In this case, the evolution equation for the metric is read
  of from
  \begin{align}\label{RicciToTFRicci}
    R_{ab}=2\Phi_{ab} + \frac{1}{4}Rg_{ab}.
  \end{align}
  For convenience, we employ such a splitting, and will use the
  trace-free Ricci tensor rather than the Schouten tensor as a dynamical
  variable.
  }
\end{IndRemark}

\medskip
There are different hyperbolic reduction strategies to turn the
tensorial expressions \eqref{CFE_tensor_zeroquants} into partial
differential equations.  In the next section, we revisit a hyperbolic
reduction that is most appropriate for the aims of this article. 

\begin{IndRemark}
  \emph{
A reader familiar with scalar-tensor and $f(R)$ theories (the two being
related for certain choices of scalar potential) might draw parallels
between the physical and unphysical variables we employ here and the
conformally related Einstein and Jordan frames (frames referring to
variable choices) in those theories---see 
\cite{Faraoni1998qx,Capozziello2011et,Clifton2011jh}, for reviews on the
topic. Although in the present case, we are not studying a modified
theory of gravity but rather a conformal version of standard general
relativity, it would be interesting to consider how one might relate the
techniques employed in modified gravity with the approach explored
here---in essence, we are treating the curvature as an independent
dynamical variable, resembling the procedure for transforming between
$f(R)$ and scalar-tensor theory. Of course, one difficulty with such a
program is the absence of a variational principle directly yielding the
CEFEs (ideally one in which the CEFE variables are treated as
independent); this will be explored this in future work.}
\end{IndRemark}

\subsection{Geometric wave equations}
\label{sec:nonlinearWaveeqs}

A first derivation of the metric conformal Einstein field equations as a
set of wave equations was given in \cite{Pae13} for the vacuum case and
in \cite{CarHurVal} for the case of trace-free matter. Our discussion
differs in the use of the trace-free Ricci tensor instead of the Schouten
tensor; we do this to make transparent the split between gauge and
non-gauge quantities. Since we only make a simple variable change, the
derivation will not be repeated here. The geometric wave equations read
\begin{subequations}\label{geo_wave}
\begin{align}
  & \square \Xi  = 4 s - \tfrac{1}{6} \Xi R ,\\
  & \square s = \Xi  \Phi _{ab} \Phi ^{ab}
  - \tfrac{1}{6} \nabla _{a}R \nabla ^{a}\Xi
  - \tfrac{1}{6} s R + \big(\tfrac{R}{12}\big)^2 \Xi
  + \tfrac{1}{4} \Xi ^5 T_{ab} T^{ab}
  - \Xi ^3 T^{ab} \Phi _{ab}
  + \Xi  T_{ab} \nabla ^{a}\Xi  \nabla ^{b}\Xi,
  \\ & \square \Phi_{ab} = 4 \Phi _{a}{}^{c} \Phi _{bc}
  - \Phi _{cd} \Phi ^{cd} g_{ab}
  -2 \Xi  \Phi ^{cd} d_{acbd} + \tfrac{1}{3} \Phi _{ab} R
  + \tfrac{1}{6} \nabla _{b}\nabla _{a}R
  - \tfrac{1}{24} g_{ab} \square R \nonumber
  \\  & \qquad \qquad \qquad \qquad \qquad \qquad
  \qquad \qquad \qquad \quad
  +   \tfrac{1}{2} \Xi ^3 T^{dc} d_{adbc}
  - \Xi  \nabla _{c}T_{b}{}^{c}{}_{a}
  - 2T_{(a|c|b)} \nabla ^{c}\Xi,
  \\ & \square d_{abcd} =  \tfrac{1}{2} d_{abcd} R
  + 2 \Xi  d_{a}{}^{e}{}_{d}{}^{p} d_{becp}
  -2 \Xi  d_{a}{}^{e}{}_{c}{}^{p} d_{bedp}
  -2 \Xi  d_{ab}{}^{ep} d_{cedp} 
  \nonumber \\ & \qquad \quad - \Xi^2T_{[c}{}^{e}d_{d]eab}
  - \Xi^2T_{[a}{}^{e}d_{b]ecd}
  + \Xi^2T^{ep}( d_{aep[c}g_{d]b} - d_{bep[c}g_{d]a} )
  + 2 \nabla_{[a}T_{|cd|b]}  \nonumber \\ &
    \qquad \qquad \qquad \qquad \qquad \qquad \qquad \qquad \qquad
    +2 \nabla_{[c}T_{|ab|d]}+ 2 g_{d[b}\nabla_{|e}T_{c|}{}^{e}{}_{a]}
    + 2 g_{c[a}\nabla_{|e}T_{d|}{}^{e}{}_{b]},
\end{align}
\end{subequations}
where $\square=\nabla_a\nabla^a$ denotes the \emph{geometric wave
operator}. Observe that these are tensorial in the same way as
expressions \eqref{CFE_tensor_zeroquants} and they do not include an
equation for the metric.  When recast in this form, the original set
of equations \eqref{CFE_tensor_zeroquants} form constraints on initial
data; the propagation of the constraints was shown in
\cite{CarHurVal}.  A further discussion of these wave equations as an
explicit system of second order hyperbolic PDEs is given in Appendix
\ref{Appendix:B} ---see also \cite{CarHurVal}.

\begin{IndRemark}
   \emph{
  To obtain a set of equations for the matter fields encoded in $T_{ab}$
  and $T_{abc}$, an explicit matter model is required. Namely, given a
  matter model consisting on some fields $\bm\tau=\{\bm\tau_1, ...
  ,\bm\tau_n\}$ so that $T_{ab}=T_{ab}(\bm\tau)$ ---and hence
  $T_{abc}=T_{abc}(\bm\tau, \nabla \bm\tau)$ using equation
  \eqref{eq:CottonExpression}---one has to derive wave equations for
  each of the fields encoded in $\bm\tau$ and their derivatives. To give
  a concise and general discussion, the wave equations for the matter
  fields will not be presented as this is a case-dependent analysis and
  can be revisited for a number of trace-free matter models in
  \cite{CarHurVal}.
  }
\end{IndRemark}

\subsection{Wave equation for metric}

The wave equation for the metric is derived from the expression of the
trace-free Ricci tensor in some fiduciary coordinate system $x^{\mu}$.
Here and in what follows, $\Gamma^{\mu}{}_{\alpha\beta}$ will denote the
Christoffel symbols of the Levi-Civita connection $\nabla$ of $\bmg$ in
the coordinate basis $x^\mu$.  First, one defines the GHG-constraint as
\begin{align}
C^{\mu} := \Gamma^{\mu} + \mathcal{H}^{\mu},
\end{align}
where $\Gamma^{\mu}:=g^{\alpha\beta}\Gamma^{\mu}{}_{\alpha\beta}$ are
the contracted Christoffel symbols and $\mathcal{H}^{\mu}$ are the
\emph{coordinate gauge source functions}.  Recall that $\Gamma^{\mu} =
- \nabla_{\alpha}\nabla^{\alpha}x^\mu$ so that setting $C^{\mu}=0$ is
equivalent to imposing the generalised harmonic gauge condition
\begin{align}\label{GHG_coordinate_Const}
\nabla_{\alpha}\nabla^{\alpha}x^\mu= \mathcal{H}^\mu.
\end{align}
with the above definitions, the \emph{reduced Ricci tensor}
$\mathcal{R}_{\mu\nu}$ is defined as
\begin{align}
\mathcal{R}_{\mu\nu}:=R_{\mu\nu} - \nabla_{(\mu}C_{\nu)}.
\end{align}
A customary calculation ---see for instance \cite{Cho08, Fri96, Fri85,
LinSchKid05, GasHil18}--- shows that the reduced Ricci tensor can be
expressed in terms of derivatives of the metric as
\begin{align}
  \mathcal{R} _{\mu \nu } = - \tfrac{1}{2} \reducedbox g_{\mu \nu } -
  \nabla _{(\mu }\mathcal{H}_{\nu) } + g^{\sigma\delta }g^{\alpha \beta}
  (\Gamma_{\nu \delta \beta }\Gamma_{\sigma \mu \alpha } +
  2\Gamma_{\delta \nu \beta }\Gamma_{(\mu\nu) \alpha } )
 \label{eq:RicciSecondDerivativesMetric}
\end{align}
where $\reducedbox:= g^{\alpha\beta}\partial_\alpha\partial_\beta$ is
the \emph{standard reduced wave operator}.  Similarly, recalling that
the Ricci scalar of $\bmg$ encodes the conformal gauge freedom, one
introduces
\begin{align}
C := R - F
\end{align}
where $F$ is the \emph{conformal gauge source function}.  Imposing the
constraint $C=0$ is equivalent to choosing a representative from the
conformal class $[\bmg]$ in the same way that imposing $C^\mu=0$ is
equivalent to choosing the coordinates to satisfy equation
\eqref{GHG_coordinate_Const} ---see Remark~\ref{remarkConfGauge}.
Therefore, using equation \eqref{RicciToTFRicci} and imposing the
GHG-coordinate and conformal gauge constraints
\[
C^\mu=0, \qquad C=0,
\]
we obtain the following reduced wave equation for the components of the
unphysical metric in the $x^{\mu}$ coordinates:
\begin{align} \label{reduced_wave_metric}
\reducedbox g_{\mu \nu } = - 4\Phi_{\mu\nu} - \frac{1}{2}F g_{\mu\nu} - 2
\nabla _{(\mu }\mathcal{H}_{\nu) } - g^{\sigma\delta }g^{\alpha \beta}
(\Gamma_{\nu \delta \beta }\Gamma_{\sigma \mu \alpha } +
2\Gamma_{\delta \nu \beta }\Gamma_{(\mu\nu) \alpha } ).
\end{align}
One can perform a similar procedure on the remaining wave equations
\eqref{geo_wave} to obtain them in their reduced form, but since it is
not absolutely necessary for the discussion that follows, that
discussion is provided in Appendix \ref{Appendix:B}.

\section{ The Conformal Einstein field equations in the linear 
  approximation }
\label{sec:lin_wave_CEFE_gen}

In this section the linearisation of the wave equations \eqref{geo_wave}
is obtained. To set up the notation regarding linearisation, first, we
outline the general procedure. Consider a one-parameter family of fields
$\bm\phi(\varepsilon)$ which satisfy an equation
\begin{align}\label{AbsModelEq}
  \mathcal{E}\bm\phi(\varepsilon)=0,
\end{align}
where $\mathcal{E}$ is some (non-linear) differential operator.  Then,
the field $\mathring{\bm\phi}:=\bm\phi(0)$ will be called the background
solution as it satisfies the equation $\mathcal{E}\mathring{\bm\phi}=0$.
Let $D$ denote the linearisation operator:
\begin{align}
  D(\mathcal{E}\mathring{\bm\phi}):=\frac{d}{d\varepsilon}(\mathcal{E}
  \mathring{\bm\phi}(\varepsilon))\Big|_{\varepsilon=0}
\end{align}
Then, using that  $D\mathring{\bm\phi}=0$, one obtains
$\mathcal{L}\delta\bm\phi=0$ where $\mathcal{L}$ is a linear operator
acting on $\delta\bm\phi:=
\frac{d\bm\phi}{d\varepsilon}|_{\varepsilon=0}$. Issues of linearisation
stability (the existence and correspondence between exact and linearised
solutions) will not be addressed here.

\medskip

For the ongoing discussion $\bm\phi = (\bm\varphi, \bmT)$, where
$\bm\varphi$ encodes all the geometric variables while $\bmf$ and $\bmT$
respectively encode the gauge and matter variables. Namely, for the
geometric sector $\bm\varphi$, one is considering an approximate
solution of the form
\begin{align}
  g_{\mu\nu}=\mathring{g}_{\mu\nu} + \delta g_{\mu\nu}, \qquad
  \Phi_{\mu\nu}= \mathring{\Phi}_{\mu\nu} + \delta \Phi_{\mu\nu},
  \qquad \Xi = \mathring{\Xi} + \delta \Xi, \qquad s= \mathring{s} +
  \delta s
\end{align}
For the gauge sector $\bmf$ one has the split
\begin{align}
  F=\mathring{F} + \delta F, \qquad 
  \mathcal{H}_{\mu} = \mathring{\mathcal{H}}_{\mu} 
  + \delta \mathcal{H}_{\mu}.
\end{align}
For the matter sector $\bmT$ one has the split
\begin{align}
  T_{\mu\nu}=\mathring{T}_{\mu\nu} + \delta T_{\mu\nu}, \qquad 
  T_{\mu\nu\alpha} = \mathring{T}_{\mu\nu\alpha} 
    + \delta T_{\mu\nu\alpha}.
\end{align}
From this point forward (unless otherwise stated), indices will be
raised and lowered with the background metric $\mathring{g}_{\mu\nu}$.

\subsection{Equation for the metric perturbation}
\label{sec:metricpert}

One of the advantages of the wave formulation of the CEFEs is that the
metric sector of the equations resembles conventional formulations of
General Relativity. This can be seen clearly even in the non-linear
equation \eqref{reduced_wave_metric} where this equation is formally
identical to the standard (non-conformal) Einstein field equations in
generalised harmonic gauge with a source term in this case given by
$-4\Phi_{ab}-\frac{1}{2}Fg_{ab}$ which could be thought conceptually as
some artificial ``geometric matter term''. To exploit the latter
viewpoint let us define
\begin{align}\label{Ssourcedef}
\mathcal{W}_{ab}:=2\Phi_{ab} + \frac{1}{4}Fg_{ab},
\end{align}
so that equation \eqref{RicciToTFRicci} agnostically reads:
\begin{align}\label{AgnosticEq}
  R_{ab}=\mathcal{W}_{ab}.
\end{align}
Expressing the equation for the metric in this way is advantageous since
one can follow the classical discussion for linearising the Einstein
field equations with respect to a general background $\mathring{g}_{ab}$
---see for instance Sec. 7.5 of \cite{Wal84}.  Since a few gauge
transformations are needed and we want to reserve the symbol $\delta
g_{\mu\nu}$ for the last transformation, we first decompose the metric
as $g_{\mu\nu}=\mathring{g}_{\mu\nu} + h_{\mu\nu}$. Then, a
straightforward linearisation of equation \eqref{AgnosticEq} gives
\begin{align}\label{step1_linearisation_metric_pert}
  \mathring{\square}{} h_{\mu \nu } =
  - 2 \delta \mathcal{W}_{\mu\nu} 
  + \dotg^{\alpha \beta } \dotnabla _{\beta}
    \dotnabla _{\mu }h{}_{\nu \alpha } 
  + \dotg^{\alpha \beta }\dotnabla _{\beta }
    \dotnabla _{\nu }h{}_{\mu \alpha } 
  - \dotg^{\alpha \beta } \dotnabla _{\mu }
    \dotnabla _{\nu }h{}_{\alpha \beta }
\end{align}
where $ \mathring{\square} :=
\mathring{\nabla}_\mu\mathring{\nabla}^\mu$ with $\mathring{\nabla}$
denoting the Levi-Civita connection of $\mathring{\bmg}$. Then, upon
commuting covariant derivatives and defining the trace-reversed metric
perturbation $\hat{h}_{\mu\nu}:=h_{\mu\nu}-\frac{1}{2}
h_{\alpha}{}^{\alpha}\dotg_{\mu\nu}$, we obtain
\begin{flalign} \label{back_wave_hcheck}
  \mathring{\square }\hat{h} {}_{\mu \nu } & = -2 \delta
  \mathcal{W}_{\mu \nu } + \delta \mathcal{W}_{\alpha}{}^{\alpha}
  \mathring{g}_{\mu \nu } -2 \mathring{R}_{\mu \alpha \nu \beta }
  \hat{h}^{\alpha \beta } + 2 \mathring{R}_{(\nu }{}^{\alpha }
  \hat{h}_{\mu) \alpha } - \mathring{g}_{\mu \nu } \dotnabla _{\beta
  }\dotnabla _{\alpha } \hat{h}^{\alpha \beta } + 2\dotnabla _{(\mu
  }\dotnabla _{|\alpha| }\hat{h}_{\nu) }{}^{\alpha }
\end{flalign}
where $\delta \mathcal{W}_{\alpha}{}^{\alpha}:=\dotg^{\alpha\beta}\delta
\mathcal{W}_{\alpha \beta}$, and $\mathring{R}_{\mu\nu\alpha\beta}$ and
$\mathring{R}_{\mu\nu}$ are the Riemann and Ricci tensors of
$\mathring{\bmg}$, which can be written in terms of the rescaled Weyl
tensor and trace-free Ricci tensors via the decomposition:
\begin{align}
  \mathring{R}_{\mu  \nu  \alpha  \beta  } = 
  \mathring{\Xi} \mathring{d}_{\mu  \nu  \alpha  \beta  }
   -2 \mathring{\Phi} _{ \beta  [\mu } \mathring{g}_{\nu] \alpha  }
   +2 \mathring{\Phi} _{  \alpha [\mu } \mathring{g}_{\nu]  \beta  }
   + \tfrac{1}{12} \mathring{R} (\mathring{g}_{\mu  \alpha  }
   \mathring{g}_{\nu  \beta  }
   - \mathring{g}_{\mu  \beta  } \mathring{g}_{\nu  \alpha  }).
\end{align}
To avoid unnecessarily long expressions, background curvature terms will
not be expanded out. Notice that equation \eqref{back_wave_hcheck}
contains only divergences of $\hat{h}_{\mu\nu}$ and derivatives of the
trace are absent. To obtain hyperbolic equations, the divergence terms
$\dotnabla _{\alpha }\hat{h} {}_{\nu }{}^{\alpha }$ are removed by
employing a slight generalisation of the procedure outlined in
\cite{Wal84}; one begins by recognising that the quantity $q_{\mu\nu} :=
h_{\mu\nu} + 2\dotnabla_{(\mu} \xi_{\nu)}$ is a gauge transformation of
the metric perturbation $h_{\mu\nu}$. Introduce its trace-reversed
version 
$\delta g_{\mu\nu}=\hat{q}_{\mu\nu}:=
q_{\mu\nu}-\tfrac{1}{2}q_{c}{}^{c}\mathring{g}_{\mu\nu}$.
Since the vector field $\xi_{\nu}$ is arbitrary, one can choose a gauge
in which $\xi_{\nu}$ satisfies an inhomogeneous wave equation. In
particular, the divergence of $\delta g_{\mu\nu}$:
\begin{align}\label{generalisedLorenzGauge}
  \dotnabla_\nu \delta g_{\mu}{}^\nu =  F_\mu,
\end{align}
may be rewritten as the following wave equation:
\begin{align}
  \mathring{\square} \xi_\mu + \mathring{R}_{\mu\nu}\xi^\nu +
  \dotnabla_\nu \hat{h}_{\mu}{}^{\nu} =  F_\mu ,
\end{align}
where $F_\mu$ is a given set of functions of the coordinates
$x^\alpha$ and the evolved fields (but not their derivatives), which
we call the {\em Lorenz gauge source functions}.  These are related to
the linearisation of the coordinate gauge source functions
$\mathcal{H}^\mu$ as discussed in Remark \ref{remark:HGSF}.  If
$\xi_{\nu}$ satisfies the above wave equation, then $\delta
g_{\mu\nu}$ satisfies the gauge \eqref{generalisedLorenzGauge}. The
gauge condition in \eqref{generalisedLorenzGauge} is a linear analogue
of the generalised harmonic gauge condition, and as such we called it
the \emph{generalised Lorenz gauge} ---a similar definition is given
also in \cite{Rainho_master_thesis} in the non-conformal context
assuming a flat background.  Working in the generalised Lorenz gauge,
we obtain the following equation for the metric perturbation:
\begin{flalign}\label{eq:metric_pert_glg}
  \mathring{\square} \delta g_{\mu \nu } & = 
  - 2 \delta \mathcal{W}_{\mu  \nu  } 
  + \delta \mathcal{W}^{\alpha  }{}_{\alpha  }
  \mathring{g}_{\mu  \nu  }
  - 2 \mathring{R}_{\mu  \alpha  \nu  \beta  } 
      \delta g^{\alpha  \beta  }
  + 2 \mathring{R}{}^{\alpha  }{}_{(\nu } \delta g_{\mu)  \alpha  }
  - \dotg_{\mu  \nu  } \dotnabla _{\alpha  }F^{\alpha  }
  + 2 \dotnabla _{(\mu  }F_{\nu)  } 
\end{flalign}
subject to the gauge condition \eqref{generalisedLorenzGauge}.

\begin{IndRemark} \label{remark:HGSF}
  \emph{The analogy between the Lorenz and Harmonic gauge is more subtle
  when dealing with a non-flat background since the direct linearisation
  of the generalised harmonic condition $C^\mu:=\Gamma^\mu +
  \mathcal{H}^\mu=0$ renders
  \begin{align}\label{linCoordGHGconst}
    \delta C^\mu=
  \dotnabla_\beta \delta g^{\beta \mu} 
    - \mathring{\mathcal{H}}^\alpha\delta g^\mu{}_{\alpha} 
    - \mathring{\Gamma}^{\mu}{}_{\alpha\beta}\delta g^{\alpha\beta} + 
     \delta \mathcal{H}^\mu =0.
  \end{align}
  Comparing equations \eqref{linCoordGHGconst} and
  \eqref{generalisedLorenzGauge} one concludes that :
  \begin{align}\label{deltFdeltHRel}
   F^\mu = \mathring{\mathcal{H}}^\alpha\delta g^\mu{}_{\alpha} 
    + \mathring{\Gamma}^{\mu}{}_{\alpha\beta}\delta g^{\alpha\beta} 
    - \delta \mathcal{H}^\mu .
  \end{align}
  Observe that even for the harmonic case $\mathring{\mathcal{H}}^\mu =
  \delta \mathcal{H}^\mu=0$ equation \eqref{linCoordGHGconst} reduces to
  \begin{align}
  \dotnabla_\beta \delta g^{\beta \mu} =
    \mathring{\Gamma}^{\mu}{}_{\alpha\beta}\delta g^{\alpha\beta},
  \end{align}
  retrieving the Lorenz gauge condition only if the background
  connection vanishes $\mathring{\Gamma}^{\mu}{}_{\alpha\beta}=0$.
  }
\end{IndRemark}

To complete the discussion now we look at the expression for $\delta
\mathcal{W}_{\mu\nu}$.  It follows from the definition of
$\mathcal{W}_{ab}$ in equation \eqref{Ssourcedef} that
\begin{align}\label{deltaSexplicit}
  \delta \mathcal{W}_{\mu \nu } = 2 \delta \Phi _{\mu \nu } 
  + \tfrac{1}{4} \mathring{F}
  \delta g_{\mu \nu } 
  - \tfrac{1}{8} \mathring{F} 
  \delta g^{\alpha}{}_{\alpha } \dotg_{\mu \nu } 
  +  \tfrac{1}{4} \delta F \dotg_{\mu \nu }.
\end{align}
Recalling that the conformal gauge constraint has been imposed $C:= R -
F=0$, observe that the linearisation of this condition leads to the
following constraint:
\begin{align}
  \delta C = 
  \delta \mathcal{W}^{\mu }{}_{\mu } 
  - \mathring{R}^{\mu \nu } \delta g_{\mu \nu } 
  + \tfrac{1}{2} \mathring{R} \delta g^{\mu }{}_{\mu } 
  - \delta F =0.
\end{align}
For consistency, the background conformal gauge source function and the
background Ricci scalar should be identified:
$\mathring{R}=\mathring{F}$.  Then, with this identification and using
equations \eqref{deltaSexplicit} the linearisation of the conformal
constraint reads
\begin{align}\label{fixtracePhipert}
\delta C = 2\delta \Phi_{\mu}{}^{\mu} -2 \mathring{\Phi}^{\mu\nu}
\delta g_{\mu\nu}=0.
\end{align}
The latter equation states that although $\Phi_{\mu\nu}$ is trace-free,
in general, its perturbation may not be. Nonetheless, the trace of the
perturbation is fixed by equation \eqref{fixtracePhipert}.  In
particular, notice that if the background is flat, this condition
reduces to $\delta \Phi_{\mu}{}^{\mu}=0$.

To obtain the final expression for the metric perturbation equation is
then enough to substitute equation \eqref{deltaSexplicit} into equation
\eqref{eq:metric_pert_glg}.  Although the resulting equation for $\delta
g_{\mu\nu}$ looks more complicated than the usual expression for the
metric perturbation this is mainly because the linearisation was
performed with respect to a general background and due to the use of the
generalised Lorenz gauge. Later on, when the background is fixed the
expressions presented in this section will simplify considerably.

\subsection{Equations for the perturbation of other conformal fields}

The linearisation of the rest of the conformal fields can be obtained
systematically.  The main calculation involves obtaining a suitable
expression for the linearisation of the geometric wave operator acting
on each of the conformal fields.  To do the calculation in the scalar
sector, first notice that for a scalar field $\varphi$ one has that
\begin{align}
  D\square \varphi = \mathring{\square} \delta \varphi
  - h^{\nu \mu  } \dotnabla _{\nu}\dotnabla_{\mu }\mathring{\varphi}
  - \dotg^{\mu    \alpha } \dotnabla _{\alpha }h{}_{\nu \mu }
  \dotnabla ^{\nu }\mathring{\varphi}
  + \tfrac{1}{2} \dotg^{\mu \alpha } \dotnabla
  _{\nu }h{}_{\mu \alpha } \dotnabla ^{\nu }\mathring{\varphi }
\end{align}
where $\mathring{\varphi}$ denotes a background quantity and $\delta
\varphi$ the perturbation. Performing the same gauge transformations of
subsection \ref{sec:metricpert} for the metric perturbation $h_{\mu\nu}$
and imposing the generalised Lorenz gauge one obtains:
\begin{align}
  D \square \varphi = \mathring{\square} \delta \varphi +
  \tfrac{1}{2} \delta g^{\mu  }{}_{\mu  } 
  \mathring{\square}\mathring{\varphi}
  - \delta g^{\mu  \nu  }  \dotnabla _{\nu  }\dotnabla _{\mu  }
  \mathring{\varphi}
  - F^{\mu  } \dotnabla _{\mu  }\mathring{\varphi}.  
\end{align}
A direct calculation using the latter expression and exploiting that the
background fields $\mathring{\bm\phi}$ satisfy the CEFEs in the form of
equations \eqref{geo_wave} as well as equations
\eqref{vanishing_CFEs_tensorial_zq}, give the following equation for the
perturbation of the conformal factor
\begin{align}\label{eq:conf_fact_pert_glg}
  \mathring{\square} \delta \Xi  =  
  - \tfrac{1}{6} \mathring{F} \delta \Xi
  - \tfrac{1}{6} \dotXi  \delta F
  + 4 \delta s 
  - \dotXi  \dotPhi ^{\mu  \nu  } \delta g_{\mu  \nu  }
  +( \tfrac{1}{24} \dotXi  \mathring{F} 
  - \mathring{s} )\delta g^{\mu  }{}_{\mu  } 
  + F^{\mu  } \dotnabla _{\mu  }\dotXi
  + \tfrac{1}{2} \dotXi ^3 \dotT^{\mu  \nu  } \delta g_{\mu  \nu  }
\end{align}
A similar calculation gives the linearised wave equation for the
Friedrich scalar. The explicit expression is lengthy and has been put in
Appendix \ref{Appendix:A}. For tensor fields the calculation is slightly
more involved. To give an abridged discussion consider now the case of
the linearisation of $\square w_\mu$, where $w_{a}$ is some covector
field.  A direct calculation gives,
\begin{align}
   D \square w_\mu& =  \mathring{\square} \delta w_{\mu }
  - \tfrac{1}{2} \mathring{w}^{\nu } \mathring{\square} h{}_{\mu \nu }  
  + \mathring{w}^{\nu } \dotnabla _{\alpha }
    \dotnabla _{[\nu }h{}_{\mu] }{}^{\alpha }
    - h{}_{\nu \alpha }
  \dotnabla ^{\alpha }\dotnabla ^{\nu }\mathring{w}_{\mu }
  + 2 \dotnabla ^{\alpha }\mathring{w}^{\nu } 
    \dotnabla _{[\nu }h{}_{\mu] \alpha }
  - \dotnabla _{\alpha }h{}_{\mu \nu } \dotnabla ^{\alpha }
    \mathring{w}^{\nu } \nonumber \\ &
  - \dotnabla ^{\nu }\mathring{w}_{\mu }
  (\dotnabla _{\alpha }h{}_{\nu }{}^{\alpha } 
  - \tfrac{1}{2}   \dotnabla _{\nu }h^{\alpha }{}_{\alpha }) 
\end{align}
where $\mathring{w}_\mu$ denotes a background quantity and $\delta
w_\mu$ the associated perturbation. Substituting the term
$\mathring{\square} h_{\mu\nu } $ using equation
\eqref{step1_linearisation_metric_pert} and rewriting the equation in
terms of the trace-reversed metric perturbation $\hat{h}_{\mu\nu}$
renders
\begin{align}\label{eq:DSqwIntermediateStep}
  D \square w_\mu & = \mathring{\square} \delta w_{\mu }
  - \mathring{w}^{\nu } \dotnabla _{\alpha   }\dotnabla _{\mu }
  \hat{h}_{\nu }{}^{\alpha } +
  \delta \mathcal{W}_{\mu }{}^{\nu } \mathring{w}_{\nu }
  - \dotnabla ^{\alpha }\mathring{w}^{\nu }( \dotnabla _{\alpha }
  \hat{h}_{\mu \nu }
  - \tfrac{1}{2} \dotg_{\mu  \nu } \dotnabla _{\alpha }
   \hat{h}^{\beta }{}_{\beta })
  \nonumber \\ &    -2 \dotnabla ^{\alpha }
  \mathring{w}^{\nu } (\dotnabla _{[\mu }\hat{h}_{\nu] \alpha } 
  - \tfrac{1}{2} g_{\alpha [\nu }  \dotnabla _{\mu] }
    \hat{h}^{\beta }{}_{\beta })
  - \dotnabla^{\alpha }\dotnabla^{\nu }
    \mathring{w}_{\mu }(\hat{h}_{\nu \alpha }
  - \tfrac{1}{2} \hat{h}^{\beta }{}_{\beta } \dotg_{\nu \alpha })
  - \dotnabla _{\alpha }\hat{h}_{\nu }{}^{\alpha } 
    \dotnabla ^{\nu }\mathring{w}_{\mu }
\end{align}
Commuting covariant derivatives on the second term in the right-hand
side of equation \eqref{eq:DSqwIntermediateStep}, performing the
gauge transformation described in subsection \ref{sec:metricpert} and
imposing the generalised Lorenz gauge one obtains
\begin{align}
  D & \square w_\mu =\mathring{\square} \delta w_{\mu }
  + \tfrac{1}{2} \delta g^{\nu }{}_{\nu } \mathring{\square}{}
    \mathring{w}_{\mu }
  - \delta g_{\nu \alpha } \dotnabla ^{\alpha }\dotnabla ^{\nu }
    \mathring{w}_{\mu }
  - \delta  g_{\nu }{}^{\alpha } \mathring{R}_{\mu \alpha } 
    \mathring{w}^{\nu }
  + \delta g^{\alpha   \beta } \mathring{R}_{\mu \alpha \nu \beta } 
    \mathring{w}^{\nu }
  - \mathcal{L}_{\bmF} \mathring{w}_\mu
  \nonumber \\ & 
  + \delta \mathcal{W}_{\mu }{}^{\nu } \mathring{w}_{\nu } 
  + 2 \dotnabla ^{\alpha }\mathring{w}^{\nu }
    ( \dotnabla _{[\nu }(\delta g)_{\mu] \alpha }
  -\tfrac{1}{2} \dotnabla _{\alpha }\delta g_{\mu   \nu }) 
  + \dotnabla^{\nu }\delta g^{\alpha }{}_{\alpha }  
    \dotnabla _{[\nu }\mathring{w}_{\mu] }
  + \tfrac{1}{2}\dotnabla_{\mu }\delta g^{\alpha }{}_{\alpha }  
    \dotnabla _{\nu }\mathring{w}^{\nu }
\end{align}
where $\mathcal{L}_{\bmF}$ denotes the Lie-derivative along $F^\mu$.
Extending the latter calculation on tensors of higher valence is a long
but straighforward calculation. The latter leads to the following
linearised equations for the conformal fields:
\begin{equation}
  \begin{aligned}
    \mathring{\square} \delta g_{\mu \nu }
    &=
    H^{g}_{(\mu\nu)}(\delta \bm\phi,
    \dotnabla\delta\bm\phi \; ; \;  \mathring{\bm\phi}, \bmf, 
    \dotnabla\mathring{\bm\phi},
    \dotnabla \bmf )
    \\
    \mathring{\square} \delta  \Xi 
    & = H^{\Xi}_{}(\delta\bm\phi, \dotnabla \delta\bm\phi
    \; ; \; 
    \mathring{\bm\phi}, \bmf, \dotnabla\mathring{\bm\phi} )  \\
    \mathring{\square} \delta  s 
    & = 
    H^{s}(\delta\bm\phi, \dotnabla\delta\bm\phi
    \; ; \; 
    \mathring{\bm\phi}, \bmf, \dotnabla\mathring{\bm\phi},
    \dotnabla \bmf )   \\ 
    \mathring{\square} \delta \Phi_{\mu \nu }
    &= 
    H^{\Phi}_{(\mu\nu)}(\delta \bm\phi,
    \dotnabla \delta \bm\phi \; ; \; 
    \mathring{\bm\phi}, \bmf, \dotnabla\mathring{\bm\phi},
      \dotnabla \bmf,  \dotnabla\dotnabla \mathring{F}, 
      \dotnabla\dotnabla \delta F )
    \\
    \mathring{\square} \delta  d_{\mu \nu \alpha \beta}
    & =
    H^{d}_{[\mu\nu][\alpha\beta]}(\delta \bm\phi, 
    \dotnabla \delta \bm\phi
    \; ; \; 
    \mathring{\bm\phi}, \bmf, \dotnabla\mathring{\bm\phi},
    \dotnabla\dotnabla\mathring{\bm\phi}, \dotnabla \bmf ) .
  \end{aligned}
\end{equation}
Here we recall that $\bm\phi$ schematically represents the conformal
fields, and the gauge sources are schematically denoted as $\bmf$;
containing the Lorenz gauge source functions $F^\mu$ as well as
$\mathring{F}$ and $\delta F$. The semicolon separates the arguments
which do not contain the evolved perturbation variables explicitly.
Observe that only the equation for $\delta \Phi_{\mu\nu}$ contains
second derivatives of $\delta F$ which restricts the allowed choices for
$\delta F$---see Remark \ref{rem:gauge_sources_ders}. The expressions
for $H^g{}_{\mu\nu}$, and $H^\Xi{}$ are given in the respective
equations \eqref{eq:metric_pert_glg} and \eqref{eq:conf_fact_pert_glg}.
The expressions for $H^s{}$, $H^\Phi_{\mu\nu}$ and
$H^{d}_{\mu\nu\alpha\beta}$ are lengthy and are given explicitly in
Appendix \ref{Appendix:A}.

\section{Analysis of wave solutions around flat spacetime}
\label{sec:Wave_CEFE_flat}

\subsection{Fixing the background solution}

The discussion given in the last section is general in the sense that
neither the background solution $\mathring{\bm\phi}$ nor the gauge
source functions $\bmf$ have been fixed.  The first observation to be
made is that, although the expression for $H^{d}_{\mu\nu\alpha\beta}$ is
complicated, in fact, if one restricts the analysis to the \emph{vacuum
case} and a \emph{conformally flat background} (so that
$\mathring{d}_{\mu\nu\alpha\beta}=0$ but $\mathring{\Phi}_{\mu\nu}\neq
0$ and $\mathring{R} \neq 0$), then the equation for the perturbation of
the rescaled Weyl tensor reduces to
\begin{align*}
  \mathring{\square} \; \delta d_{\mu\nu\alpha\beta}
  = \frac{1}{2}\mathring{F}  \delta d_{\mu\nu\alpha\beta},
\end{align*}
and this equation decouples from the rest.  This is to be expected since
as one way to study linearised gravity is through the spin-2 equation
for which the metric perturbation plays no role ---see \cite{PenRin84,
PenRin86}. Since the aim of this article is to give a first analysis
into the linearised version of the CEFEs, it is relevant to now examine
the complete system even if some of the equations decouple as in the
case described above.

\begin{figure}[t!]
  \begin{center}
     \begin{subfigure}{}
   \includegraphics[width=4cm]{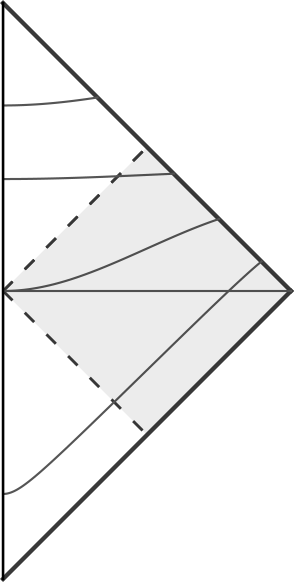}   
     \put(-120,230){\Large{$i^+$}}
     \put(-50,170){\Large{$\mathscr{I}^{+}$}}
     \put(-60,115){$t=0$}
     \put(2,110){\Large{$i^0$}}
     \put(-80,125){\rotatebox{22}{$\tau=\tau_\star$}}
     \put(-50,50){\Large{$\mathscr{I}^{-}$}}
     \put(-120,-13){\Large{$i^-$}}
     \end{subfigure}
     \hspace{2cm}
     \begin{subfigure}{}
     \includegraphics[width=7cm]{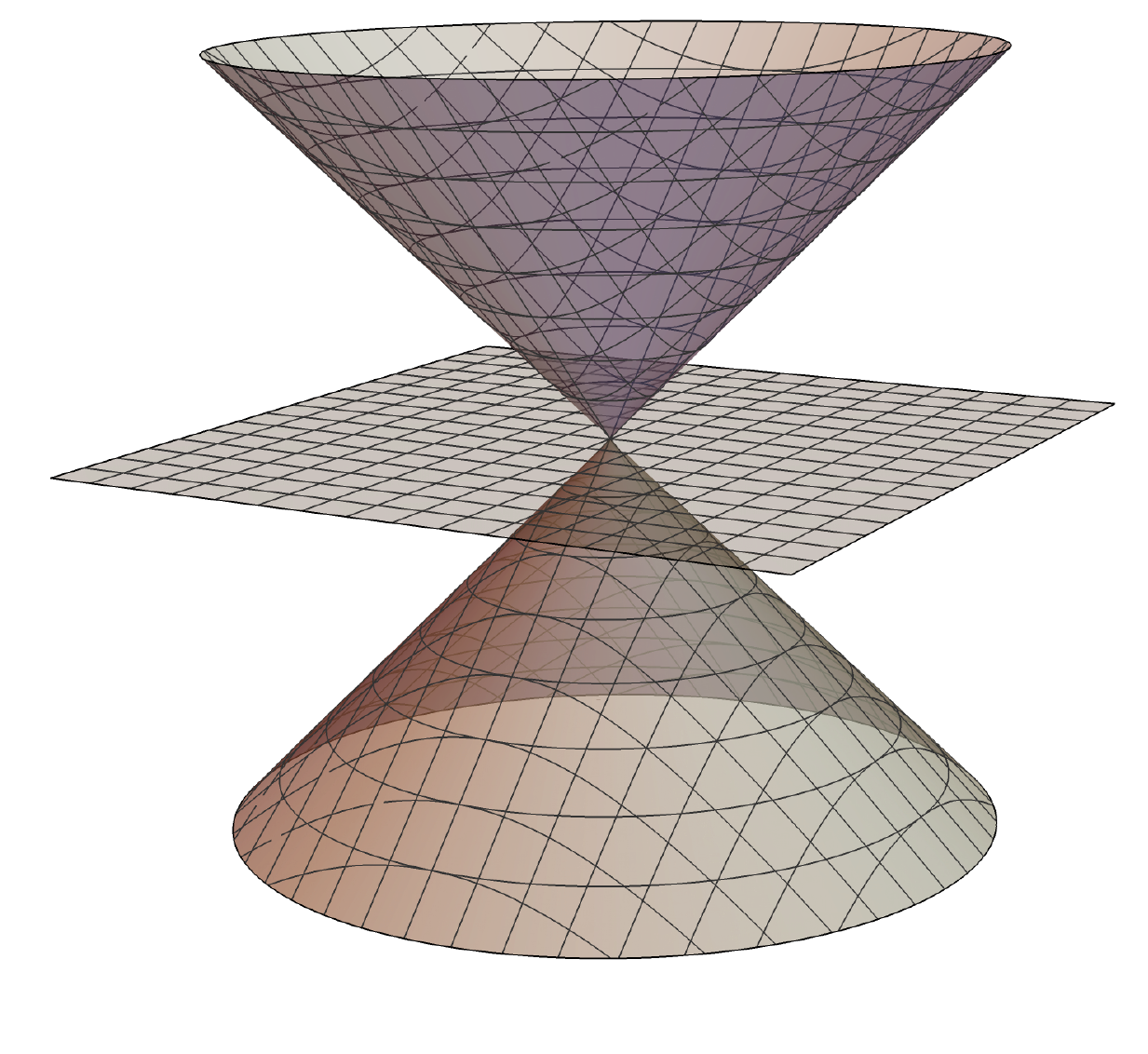}
     \put(-25,45){\Large{$\mathscr{I}^{-}$}}
     \put(-30,160){\Large{$\mathscr{I}^{+}$}}
     \put(-80,100){\LARGE{$i^0$}}
     \put(-2,108){\large{$t=0$}}
     \end{subfigure}
   \end{center}
  \caption{In the left panel a Penrose diagram of the Minkowski
    spacetime is shown where the shaded area represents the region
    covered by the Inversion representation.  The Cauchy hypersurface
    $\tilde{t}=t=0$ as well as some hyperboloidal surfaces
    $\tau=\tau_{\star}$ are also depicted. The right panel shows a
    neighbourhood of $i^0$ in the Inversion representation of the
    Minkowski spacetime. As depicted in the figure
    $i^0$ is located at $\rho=0$ and $\mathscr{I}^{\pm}$
    correspond to the lightcone through the origin in the unphysical
    spacetime.}\label{fig:diagram}
\end{figure}
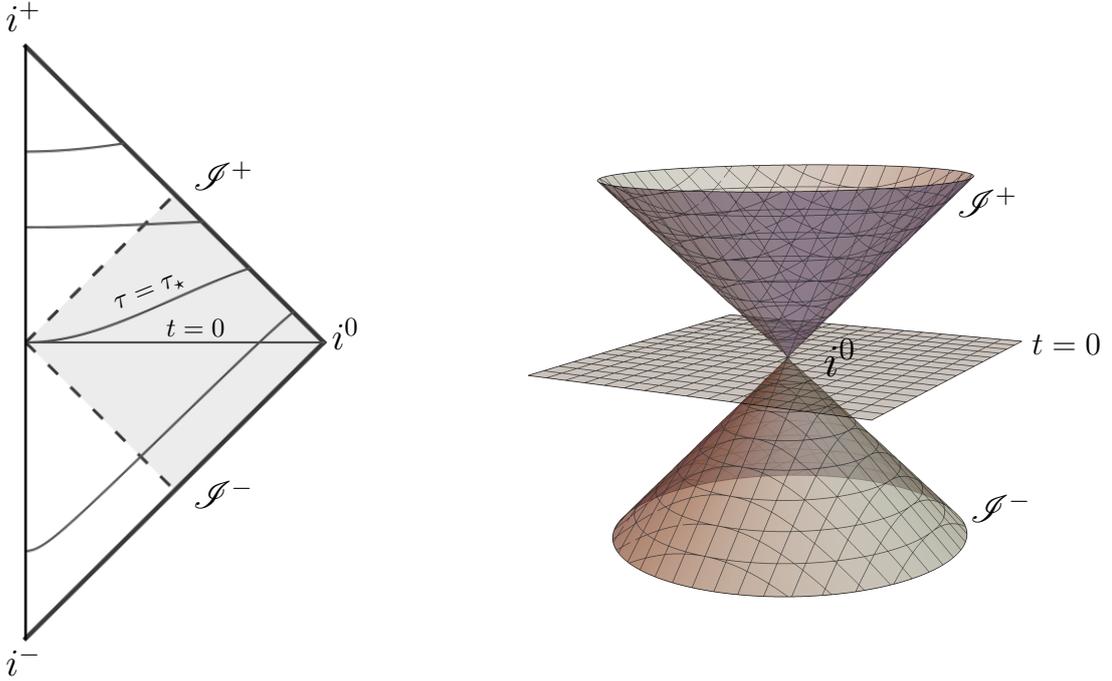

The simplest choice of background is one in which the background
geometry is flat. Naively, one would think that this forces the
background conformal factor to be constant, $\mathring{\Xi}=1$, and the
equations trivialise to the standard non-conformal case; this would
defeat the purpose of considering the CEFEs in the first place, as one
would not be able to incorporate the conformal boundary
in such an analysis. Fortunately, there
exists a non-trivial conformal transformation that maps the Minkowski
spacetime into itself such that spatial infinity and null infinity for
the \emph{physical} Minkowski spacetime are respectively mapped to the
origin and the lightcone through the origin of the (conformally
transformed) \emph{unphysical} Minkowski spacetime. The evaluation of
the field at the conformal boundary of the physical spacetime
corresponds to evaluation at finite coordinate locations in the
unphysical spacetime.  We point out that such a conformal transformation
does not necessarily yield a full compactification, as some points (such
as the physical origin) are mapped to infinity ---see Figure
\ref{fig:diagram}.  This choice for a flat background, which we call the
\emph{inversion representation of the Minkowski spacetime}, is the
simplest non-trivial case (in the sense that it can incorporate the
conformal boundary) that one can analyse in the conformal set-up.
Therefore, the following analysis can be regarded as the conformal
counterpart of the standard discussion of linearised (physical) Einstein
field equations around flat spacetime. The inversion representation of
the Minkowski spacetime is a compelling choice not only for its
simplicity, but also because it is related to the
\emph{$i^0$-cylinder representation of the Minkowski spacetime}
(see for instance \cite{Fri02, GasVal20, GasVal21a, MinMacVal22,
ValAli22,DuaFenGasHil23}) which has
$\mathring{d}_{\mu\nu\alpha\beta}=0$, $\mathring{R} = 0$ with a
non-trivial trace-free Ricci tensor $\mathring{\Phi}_{\mu\nu}\neq 0$.
While an analysis on the $i^0$-cylinder background
would be of great interest, the sources $\bmH$ in
the equations for the fields other than $\delta d_{\mu\nu\alpha\beta}$
(see Appendix \ref{Appendix:A}) become cumbersome when
$\mathring{\Phi}_{\mu\nu}\neq 0$; this will be pursued in future work.

\subsection{The inversion conformal representation of the Minkowski spacetime}
\label{sec:InversionMinkowski}

The inversion representation of the Minkowski spacetime described in the
preceding paragraph is a standard conformal representation detailed in
\cite{Ste91,GasVal21a,MinMacVal22,DuaFenGasHil23}, which we now
summarise. In what follows $(\mathbb{R}^4, \tilde{\bmeta})$ will denote
the (physical) Minkowski spacetime.  Let $\tilde{x}^{\tilde{\mu}}=
(\tilde{t},\tilde{x}^{\tilde{i}})$, represent \emph{physical} Cartesian
coordinates.  The associated coordinate basis vectors and covectors are
denoted by $\bm\partial_{\tilde{\mu}}$ and
$\mathbf{d}\tilde{x}^{\tilde{\mu}}$, respectively. The components of an
arbitrary abstract tensor $S_{ab}$ in the physical Cartesian coordinate
basis are denoted by $S_{\tilde{\mu}\tilde{\nu}}$. The Minkowski metric
reads 
\begin{align}
\tilde{\bmeta}=
\eta_{\tilde{\mu}\tilde{\nu}}\mathbf{d}\tilde{x}^{\tilde{\mu}}
\otimes
\mathbf{d}\tilde{x}^{\tilde{\nu}},
\end{align}
where~$\tilde{\eta}_{\tilde{\mu}\tilde{\nu}}=\text{diag}(-1,1,1,1)$.
Defining the \emph{physical} radial coordinate via~$\tilde{\rho}^2=
\delta_{\tilde{i}\tilde{j}}\tilde{x}^{\tilde{i}}\tilde{x}^{\tilde{j}}$
where~$\delta_{\tilde{i}\tilde{j}}=\text{diag(1,1,1)}$ and considering
an arbitrary choice of coordinates on~$\mathbb{S}^2$ the physical
Minkowski metric can be written as
\begin{align}\label{MinkowskiMetricPhysicalPolar}
\tilde{\bmeta}=-\mathbf{d}\tilde{t}\otimes\mathbf{d}\tilde{t}
+\mathbf{d}\tilde{\rho}\otimes \mathbf{d}\tilde{\rho}+\tilde{\rho}^2
\mathbf{\bm\sigma},
\end{align}
with~$\tilde{t}\in(-\infty, \infty)$, $\tilde{\rho}\in [0,\infty)$
where~$\bm\sigma$ denotes the standard metric on~$\mathbb{S}^2$. 

We now introduce the \emph{unphysical Cartesian coordinates}
$x^{\mu}=(t,x^{i})$, and denote the associated vector and covector basis
with $\bm\partial_{\mu}$ and $\mathbf{d}x^{\mu}$, respectively. As
before, the components of an arbitrary tensor $S_{ab}$ in the unphysical
Cartesian coordinate basis will be denoted as $S_{\mu\nu}$. The
relationship between the physical and unphysical Cartesian coordinates
is given by:
\begin{align}
x^{\mu}
=
\delta_{\tilde{\mu}}{}^{\mu}\;\tilde{x}^{\tilde{\mu}}/{\tilde{X}^2}, 
\qquad 
\tilde{X}^2 
\equiv
\tilde{\eta}_{\tilde{\mu}\tilde{\nu}}\tilde{x}^{\tilde{\mu}}
\tilde{x}^{\tilde{\nu}},
\label{UnphysicalCartesianToPhysicalCartesian}\\
\tilde{x}^{\tilde{\mu}}=\delta^{\tilde{\mu}}{}_{\mu}\;x^{\mu}/X^2,
\qquad
X^2=\eta_{\mu\nu}x^{\mu}x^{\nu}. 
\label{PhysicalCartesianToUnphysicalCartesian}
\end{align}
where $\delta_{\tilde{\mu}}{}^{\mu}=\text{diag}(1,1,1,1)$,
$X^2=1/\tilde{X}^2$ and $\eta_{\mu\nu}=\text{diag}(-1,1,1,1)$. This
coordinate transformation is valid in the complement of the lightcone at
the origin in the physical Minkowski spacetime where $\tilde{X}^2 > 0$.

\begin{IndRemark}
  \emph{ The notational reason for having a tilde over the physical
  coordinate indices is to distinguish the physical and unphysical
  coordinate bases denoted by $\bm\partial_\mu$ and
  $\bm\partial_{\tilde{\mu}}$.  The latter requires the use of
  $\delta_{\tilde{\mu}}{}^{\mu}$ in the expressions
  \eqref{UnphysicalCartesianToPhysicalCartesian} to keep the tilded and
  untilded indices balanced. This notation is inspired by that of the
  dual foliation formalism of \cite{Hil15}.
  }
\end{IndRemark}

%
%

From the relation between the physical and the (unphysical) inversion
coordinate bases one can construct the Jacobians
\begin{align}\label{eq:Jacobians}
   J_{\tilde{\mu}}{}^{\mu}:= 
    \frac{\partial x^\mu}{\partial \tilde{x}^{\tilde{\mu}}}
    =
    \tilde{X}^{-2}(\delta_{\tilde{\mu}}{}^{\mu}
    - 2 x^\mu\tilde{x}_{\tilde{\mu}}), \qquad 
   J_{\mu}{}^{\tilde{\mu}}{}:=
   \frac{\partial \tilde{x}^{\tilde{\mu}}}{\partial x^\mu}=
    X^{-2}
      (\delta^{\tilde{\mu}}{}_{\mu}-2\tilde{x}^{\tilde{\mu}}x_{\mu})
\end{align}
To identify a suitable conformal metric, it is enough to compute
$ \eta_{\mu\nu}\mathbf{d}x^{\mu}\otimes \mathbf{d}x^{\nu} =
 \eta_{\mu\nu}
  J_{\tilde{\mu}}{}^{\mu}J_{\tilde{\nu}}{}^{\nu}
  \mathbf{d}\tilde{x}^{\tilde{\mu}}
  \otimes
  \mathbf{d}\tilde{x}^{\tilde{\nu}}
  $.
Introducing  $\bm\eta=\eta_{\mu\nu}\mathbf{d}x^{\mu}\otimes
\mathbf{d}x^{\nu}$, we write this compactly as:
\begin{align}
\label{InverseMinkowskiMetricDef}
  \bm\eta=\Xi^2 \hspace{0.5mm}\tilde{\bmeta},
\end{align}
with $\Xi =X^2$. This demonstrates that the inversion (unphysical)
Minkowski spacetime can be recast as a conformal transformation of the
physical Minkowski spacetime.  Although, in most of the upcoming
discussion, all tensor components will be expressed in just one
coordinate basis (the unphysical one), notice that, the relation
between the components of a tensor $S_{ab}$ in the physical and the
unphysical bases is then
given by
\begin{align}\label{CompsRelation}
  S_{\tilde{\mu}\tilde{\nu}}= \Xi^2(\delta_{\tilde{\mu}}{}^\mu 
  - 2x^\mu\tilde{x}_{\tilde{\mu}})
  (\delta_{\tilde{\nu}}{}^\nu -2x^\nu\tilde{x}_{\tilde{\nu}})S_{\mu\nu}.
\end{align}
To complete the discussion, one can construct an unphysical spherical
polar coordinate system. Upon introducing
$\rho^2=\delta_{ij}x^{i}x^{j}$, a calculation shows that the unphysical
 (conformal) metric~$\bm\eta$ and conformal factor~$\Xi$ read
\begin{align}
\label{InverseMinkowskiUnphysicaltrhocoords}
\bm\eta=-\mathbf{d}t\otimes\mathbf{d}t +\mathbf{d}\rho\otimes
\mathbf{d}\rho+\rho^2 \mathbf{\bm\sigma}, \qquad \Xi=\rho^2 -t^2,
\end{align}
with~$t\in(-\infty,\infty)$ and~$\rho\in~[0,\infty)$.  Notice that
spatial infinity $i^0$ of the physical Minkowski spacetime
$(\mathbb{R}^4, \tilde{\bmeta})$ is mapped to the origin $(t=0,\rho=0)$
in $(\mathbb{R}^4,\bm\eta)$. Also, future and past null infinity
$\mathscr{I}^\pm$ of the physical Minkowski spacetime are mapped to the
lightcone passing through the origin. In other words, introducing the
unphysical retarded and advanced times $u:=t-\rho$ and $v:=t+\rho$,
future/past null infinity $\mathscr{I}^\pm$ is located at $v=0$ and
$u=0$ respectively.  To round up the discussion between the relation the
physical and unphysical coordinates, here we record that
\begin{align}
 \label{physicalToUnphysicaltrho}
   t=\frac{\tilde{t}}{\tilde{\rho}^2-\tilde{t}^2}, \qquad \rho =
   \frac{\tilde{\rho}}{\tilde{\rho}^2-\tilde{t}^2}.
\end{align}


\subsection{Linearised equations around the Minkowski-inversion 
            background}

\subsubsection{The background solution}

We now fix the background solution $\mathring{\bm\phi}$ to be the
Minkowski inversion background discussed in the previous subsection,
\ref{sec:InversionMinkowski}. This yields the following expressions for
the background fields
\begin{align}\label{eq:backgroundsolution}
  \mathring{g}_{\mu\nu}= \eta_{\mu\nu}, \qquad \mathring{\Xi}=
  \rho^2-t^2, \qquad \mathring{s}= 2, \qquad \mathring{T}_{\mu\nu}=0, 
  \qquad \mathring{T}_{\mu\nu\alpha}=0.
\end{align}
the latter implies that the curvature
$\mathring{R}_{\mu\nu\alpha\beta}=0$ vanishes which, in terms of its
irreducible decomposition implies that
\begin{align*}
\mathring{\Phi}_{\mu\nu}=0, \qquad \mathring{R}=0,
\qquad \mathring{d}_{\mu\nu\alpha\beta}=0.
\end{align*}
Since one has that $\mathring{R}=\mathring{F}$, this fixes the
background conformal gauge source function. Nonetheless, the remaining
gauge source functions encoded in $F^\mu$ and $\delta F$ need not
vanish and will (along with matter perturbation quantities) be left
unspecified. This gives the following equations
\begin{subequations}\label{linearWaveEqs}
\begin{align}
  \mathring{\square} \delta g _{\mu  \nu  } & = 
  - 4 \delta \Phi _{\mu  \nu  }
  + 2 \delta \Phi ^{\alpha  }{}_{\alpha  } \mathring{g}_{\mu  \nu  }
  - \mathring{g}_{\mu  \nu  } \dotnabla _{\alpha  }F^{\alpha  }
  + 2\dotnabla _{(\mu  }F_{\nu)  }
  + \tfrac{1}{2}\delta F \dotg_{\mu\nu}
  \label{linMetric}
  \\
  \mathring{\square} \delta \Xi  & = 4 \delta s  
  - \mathring{s} \delta g ^{\mu  }{}_{\mu  } 
  + F^{\mu}\dotnabla_\mu \mathring{\Xi} - \tfrac{1}{6}\dotXi \delta F\\
  \mathring{\square} \delta s  & = 
  \dotXi \delta T^{\mu\nu}\dotnabla_\mu\dotXi\dotnabla_\nu\dotXi
  - \tfrac{1}{6}\mathring{s} \delta F 
  - \tfrac{1}{6}\dotnabla_\mu\delta F \dotnabla^\mu \dotXi\\
  \mathring{\square} \delta \Phi_{\mu  \nu  }
  & = -2 \delta T_{(\mu}{}^\alpha{}_{\nu)}\dotnabla_\alpha \dotXi 
  - \dotXi \dotnabla_\alpha \delta T_{\mu}{}^\alpha{}_{\nu}
  + \tfrac{1}{6}\dotnabla_{\{\mu}\dotnabla_{\nu\}} \delta F 
  \label{linPhi} \\
  \mathring{\square} \delta d_{\mu\nu\alpha\beta} =&  
  2 \dotnabla_{[\mu}\delta T_{|\alpha\beta|\nu]} 
  +2  \dotnabla_{[\alpha}\delta T_{|\mu\nu|\beta]} + 2
  \dotg_{\beta[\nu}\dotnabla_{|\sigma}
  \delta T_{\alpha|}{}^{\sigma}{}_{\mu]} 
  + 2
  \dotg_{\alpha[\mu}\dotnabla_{|\sigma}
  \delta T_{\beta|}{}^{\sigma}{}_{\nu]}. \label{linRescaledWeyl}
\end{align}
\end{subequations}
\begin{IndRemark}
  \label{rem:gauge_sources_ders}
  \emph{The Lorenz gauge source functions $F^\mu$ are allowed to
  depend on the variables $\delta \bm\phi$ as they appear only with first
  derivatives in equations \eqref{linearWaveEqs}. However, the
  conformal gauge source function $\delta F$ is only allowed to depend
  on the coordinates due to the second derivatives of $\delta F$ in
  the equation \eqref{linPhi}.}
\end{IndRemark}

\begin{IndRemark}
  \emph{Although for simplicity we have fixed the notation in
    subsection \ref{sec:InversionMinkowski} so that $x^\mu$ denotes
    the unphysical Cartesian system of coordinates, in fact, equations
    \eqref{linearWaveEqs} are not bound to these coordinates, the only
    assumption used was that the background Riemann curvature
    $\mathring{R}^{\mu}{}_{\alpha \beta \nu}$ vanishes. In other
    words, in deriving equation \eqref{linearWaveEqs}, the vanishing
    of the (background) Christoffel symbols
    $\mathring{\Gamma}^\alpha{}_{\mu\nu}$ has not been assumed,
    consequently, formally identical equations to
    \eqref{linearWaveEqs} will hold in any coordinate
    system. Nonetheless, the unphysical Cartesian system $x^\mu$ will
    be assumed in the discussion of subsection \ref{sec:gen_sol_lin}
    so that one may apply standard techniques such as expressing the
    solutions in terms of Greens functions and multipolar expansion
    approaches.}
\end{IndRemark}

\subsubsection{The matter relations}
\label{sec:MatterRelations}
The matter equations depend on the particular matter model in question,
however, since the background matter terms are assumed to vanish, the
direct linearisation of the general relations, combined with the
symmetries of $T_{abc}$ and $T_{ab}$ described in section
\ref{Sec:CFEs}, yield a set of formally identical expressions that the
perturbations must satisfy:
\begin{subequations}\label{eq:LinMatterGeneralRels}
\begin{align} 
  \delta T_{\mu}{}^{\mu}=0, \qquad \dotnabla_\nu \delta T_{\mu}{}^\nu=0,
  \qquad
  \delta T_{\alpha\beta\mu}= \delta T_{[\alpha\beta]\mu},
  \qquad
  \delta T_{[\alpha\beta\mu]}=0, \qquad \dotnabla_\alpha 
  \delta T_{\mu\nu}{}^\alpha=0.
\end{align}
Also, one has the relations:
\begin{align}
   \delta T_{\alpha \beta \gamma}& := \dotXi \dotnabla_{[\alpha} 
   \delta T_{\beta]\gamma}
  + 3 \delta T_{\gamma[\beta}\dotnabla_{\alpha]}\mathring{\Xi}
  -\dotg_{\gamma[\alpha} \delta T_{\beta]\lambda}\dotnabla^\lambda\dotXi,
  \label{eq:CottToEne}
  \\ 
  \dotnabla _{\alpha  }\delta T_{\mu  }{}^{\alpha  }{}_{\nu  }
  &=- \tfrac{1}{2} \mathring{\Xi}  \mathring{\square}
  \delta T_{\mu  \nu  }
  -4 \mathring{s} \delta T_{\mu  \nu  }  
  -2 \dotnabla _{\alpha  } \delta T_{\mu  \nu  } 
  \dotnabla ^{\alpha  }\mathring{\Xi}
  + \dotnabla ^{\alpha  }\mathring{\Xi} 
  \dotnabla _{(\mu  } \delta T_{\nu)  \alpha  }.
  \label{eq:CottToEneDeriv}
\end{align}
\end{subequations}

\subsubsection{Propagation of the generalised Lorentz gauge}

Tracing equation \eqref{linPhi} and using equations
\eqref{eq:LinMatterGeneralRels} shows that
\begin{align}\label{tracelinPhi}
\mathring{\square } \delta \Phi_{\mu}{}^{\mu}=0.
\end{align}
Recall that the linearisation of the conformal gauge constraint,
translates in the present case, to the condition
\begin{align}\label{LinConfConst}
\delta \Phi_{\mu}{}^{\mu}=0.
\end{align}
This, in particular, simplifies equation \eqref{linMetric}:
\begin{align}
\mathring{\square} \delta g _{\mu  \nu  } & = 
  - 4 \delta \Phi _{\mu  \nu  }
  - \mathring{g}_{\mu  \nu  } \dotnabla _{\alpha  }F^{\alpha  }
  + \dotnabla _{\mu  }F_{\nu  } + \dotnabla _{\nu  }F_{\mu  }
  + \tfrac{1}{2}\delta F \dotg_{\mu\nu} .
  \label{linMetricSim}
\end{align}
The condition \eqref{LinConfConst} is trivially propagated by equation
\eqref{tracelinPhi}. Namely, if initial data is given such that
\begin{align*}
\delta \Phi_{\mu}{}^{\mu}=0 \qquad \text{and} \qquad
    \dotnabla_n \delta \Phi_{\mu}{}^{\mu}=0 \qquad
  \text{on} \qquad \Sigma ,
\end{align*}
where  $\dotnabla_n:=n^\mu\dotnabla_\mu$ and $n^\mu$ denotes the normal
to a spacelike hypersurface $\Sigma$. Then, by existence and uniqueness
of solutions to the flat-wave equation \eqref{tracelinPhi}, one has
\begin{align*}
\delta \Phi_{\mu}{}^{\mu}=0 \qquad \text{on} \qquad \mathcal{U} 
\subset \mathcal{D}^{+}(\Sigma).
\end{align*}
where $\mathcal{U}$ is an open set and $\mathcal{D}^{+}(\Sigma)$ denotes
the future domain of dependence of $\Sigma$. This result can be
understood as the propagation of the conformal constraint in the linear
setting. To show that generalised Lorenz gauge condition
\eqref{generalisedLorenzGauge} propagates one needs to use the
contracted Bianchi identity. In the CEFE language this is encoded in
equation \eqref{standardCEFESchouten}. Hence, for completeness and
future reference, we linearise the CEFEs as defined in equations
\eqref{CFE_tensor_zeroquants} and \eqref{vanishing_CFEs_tensorial_zq}
and record the result in Remark \ref{remark_linear_cefe}.
\begin{IndRemark}
  \label{remark_linear_cefe}
  \emph{
  Using equation equation \eqref{Ricci_ToSchouten} to express
  \eqref{CFE_tensor_zeroquants} using the trace-free Ricci tensor and
  linearising around a flat background gives
  \begin{subequations}\label{eqLinearCEFEs}
  \begin{align}
    &\dotnabla _{\mu  } \delta s  = 
    - \delta \Phi _{\mu  \nu  } \dotnabla ^{\nu  }\dotXi
    + \tfrac{1}{2} \dotXi ^2  \delta T_{\mu  \nu  } 
    \dotnabla ^{\nu  }\dotXi
    -\tfrac{1}{24}\delta F \dotnabla_\mu \dotXi, \\
    & \dotnabla _{\mu  } \delta \Phi _{\alpha  \nu  }  
    - \dotnabla _{\nu  } \delta \Phi _{\alpha  \mu  }
    = \dotXi  \delta T_{\mu  \nu  \alpha  }
    - \delta d_{\alpha  \beta  \mu  \nu  } \dotnabla ^{\beta  }\dotXi
    +  \tfrac{1}{12} \dotg_{\alpha  [\mu  } \dotnabla _{\nu]  }\delta F,
    \label{eq:CEFELinPhi} \\
    & \dotnabla ^{\mu  }\delta d_{\mu\nu \alpha  \beta  } =  
    \delta T_{\alpha  \beta  \nu  }, \label{eq:CEFElinWeyl} \\
    &\dotnabla _{\mu  }\dotnabla _{\nu  } \delta \Xi  =
    \Gamma[q]^\lambda{}_{\mu\nu}\dotnabla_\lambda \dotXi 
    +\mathring{s}q_{\mu\nu} 
    + (\delta s - \tfrac{1}{24}\dotXi\delta F)\dotg_{\mu\nu}
    \dotXi \delta \Phi_{\mu\nu} + \tfrac{1}{2}\dotXi \delta T_{\mu\nu},
    \label{eq:CEFEconf_factor} \\
    & \delta R_{\mu\nu\alpha\beta}=
    \dotXi  \delta d_{\mu  \nu  \alpha  \beta  } 
    + 2\delta \Phi _{[\nu | \beta  } \dotg_{  \alpha| \mu]  }   
    + 2\delta \Phi _{[\mu | \alpha  } \dotg_{  \beta| \nu]  }
    + \tfrac{1}{6} \delta F \dotg_{[\mu|  \alpha  } 
      \dotg_{ \beta |\nu]   },
    \label{eq:CEFE_Riem_Alg_Geo}
  \end{align}
  where
  \begin{align}
    \Gamma[q]^{\alpha}{}_{\mu\nu}  :=
    \tfrac{1}{2}\mathring{g}^{\alpha\beta}
    \Big( \dotnabla _{\mu  }  q_{\nu  \beta  }
    +  \dotnabla _{\nu  }  q_{\mu  \beta  } 
    -  \dotnabla _{\beta  } q_{\mu  \nu  }\Big) ,
    \qquad
    \delta R_{\mu  \nu  \alpha  \beta  } :=
    2\dotnabla _{[\alpha  }\dotnabla _{|[\nu  } q_{\mu  ]|\beta]  }.
  \end{align}
  \end{subequations}
  where the expressions have been written using $q_{\mu\nu}$ instead of
  $\delta g_{\mu\nu}$ to avoid long expressions. Recall that  $\delta
  g_{\mu\nu}$ is simply the trace-reversed version of $q_{\mu\nu}$.
  }
\end{IndRemark}

Contracting equation \eqref{eq:CEFELinPhi}, gives $\dotnabla_\nu \delta
\Phi_{\mu}{}^{\nu}= \dotnabla_\mu  \delta \Phi_\nu{}^\nu +
\tfrac{1}{8}\dotnabla_\mu \delta F$ and using equation
\eqref{LinConfConst} one obtains
\begin{align}\label{LinBianchiContracted}
  \dotnabla_\nu  \delta \Phi_{\mu}{}^{\nu}= \frac{1}{8}\dotnabla_\mu \delta F.
\end{align}
The divergence-fixed condition \eqref{LinBianchiContracted} is also
propagated by the wave equations \eqref{linearWaveEqs}. To see this,
notice that applying $\mathring{\nabla}^\mu$ to equation \eqref{linPhi},
using \eqref{eq:LinMatterGeneralRels}, and after trivially commuting the
flat-covariant derivative $\dotnabla$ one obtains
\begin{align}\label{eq:linearisedConfConsProp}
  \mathring{\square} (\dotnabla^\mu \delta \Phi_{\mu\nu}
  - \tfrac{1}{8}\dotnabla_\nu \delta F)=0.
\end{align}
Hence, if initial data is given on $\Sigma$ such that $\dotnabla^\mu
\delta \Phi_{\mu\nu}=\tfrac{1}{8}\dotnabla_\nu \delta F$ and
$\dotnabla_n ( \dotnabla^\mu \delta \Phi_{\mu\nu}
-\tfrac{1}{8}\dotnabla_\nu\delta F) =0$ then condition
\eqref{LinBianchiContracted} is satisfied in $\mathcal{U}\subset
\mathcal{D}^{+}(\Sigma)$. To show propagation of the generalised Lorenz
gauge \eqref{generalisedLorenzGauge}, one proceeds analogously, taking a
divergence of equation \eqref{linMetricSim} and using equations
\eqref{LinConfConst} and \eqref{LinBianchiContracted} gives the
following equation
\begin{align}\label{eq:lorenzpropmain}
\dotnabla^\mu \mathring{\square} \delta g_{\mu\nu} = 
\mathring{\square} F_{\nu}
\end{align}
which can be rewritten as
\begin{align}\label{propagationGenLorenz}
\mathring{\square} (\dotnabla^\mu \delta g_{\mu\nu} -F_\nu) = 0.
\end{align}
Hence, if one gives initial data on $\Sigma$ such that $\dotnabla^\mu
\delta g_{\mu\nu} -F_\nu=0$ and $ \dotnabla_n(\dotnabla^\mu
\delta g_{\mu\nu} -F_\nu)=0$ then by virtue of equation
\eqref{propagationGenLorenz} one has that
\eqref{generalisedLorenzGauge}, namely $\dotnabla^\mu \delta g_{\mu\nu}
=F_\nu$, is satisfied on $\mathcal{D}^{+}(\Sigma)$. In other words, the
\emph{generalised Lorenz gauge propagates}. This discussion is
summarised in the following
\begin{IndProposition}
  \emph{The linearisation of the wave-CEFEs around
  the inversion-Minkowski unphysical spacetime is given by
  \begin{subequations}\label{LinCEFEprop}
  \begin{align}
    \mathring{\square}\delta g_{\mu  \nu  } & = 
    -4 \delta \Phi _{\mu  \nu  }
    - \mathring{g}_{\mu  \nu  } \dotnabla _{\alpha  }F^{\alpha  }
    + 2\dotnabla _{(\mu  }F_{\nu ) }  +
    \tfrac{1}{2}\delta F \dotg_{\mu\nu}
    \label{linMetricClean}
    \\
    \mathring{\square} \delta \Xi  & = 4 \delta s  
  - \mathring{s} \delta g ^{\mu  }{}_{\mu  } 
  + F^{\mu}\dotnabla_\mu \mathring{\Xi} - \tfrac{1}{6}\dotXi \delta F
  \label{linCF}\\
   \mathring{\square} \delta s  & = 
  \dotXi \delta T^{\mu\nu}\dotnabla_\mu\dotXi\dotnabla_\nu\dotXi
  - \tfrac{1}{6}\mathring{s} \delta F 
  - \tfrac{1}{6}\dotnabla_\mu\delta F \dotnabla^\mu \dotXi
  \label{linFscalar} \\
  \mathring{\square} \delta \Phi_{\mu  \nu  }
  & = -2 \delta T_{(\mu}{}^\alpha{}_{\nu)}\dotnabla_\alpha \dotXi 
  - \dotXi \dotnabla_\alpha \delta T_{\mu}{}^\alpha{}_{\nu}
  + \tfrac{1}{6}\dotnabla_{\{\mu}\dotnabla_{\nu\}} \delta F
  \label{linTFRicci} \\
    \mathring{\square} \delta d_{\mu\nu\alpha\beta} =&
    2  \dotnabla_{[\mu}\delta T_{|\alpha\beta|\nu]} 
    + 2  \dotnabla_{[\alpha}\delta T_{|\mu\nu|\beta]} 
    + 2 \dotg_{\beta[\nu}\dotnabla_{|\sigma}
        \delta T_{\alpha|}{}^{\sigma}{}_{\mu]} 
    + 2 \dotg_{\alpha[\mu}\dotnabla_{|\sigma}
        \delta T_{\beta|}{}^{\sigma}{}_{\nu]}. \label{linResWeyl}
  \end{align}
  \end{subequations}
  where the matter sources satisfy the relations in equation
  \eqref{eq:LinMatterGeneralRels}. The conditions $\delta
  \Phi_{\mu}{}^{\mu}=0$, and $\dotnabla^{\mu}\delta
  \Phi_{\mu\nu}=\frac{1}{8}\dotnabla_\mu\delta F$, along with the
  generalised Lorenz gauge condition $\dotnabla^{\mu}\delta
  g_{\mu\nu}=F_\mu$, propagate.
  }
\end{IndProposition}
%
\begin{IndRemark}
  \emph{
  Observe that if $\delta F=0$ is set then $\delta \Phi_{\mu\nu}$ is a
  \emph{TT-tensor} (transverse and tracefree).
  Furthermore, if the Lorenz gauge condition is chosen, $F_{\mu}=0$,
  equation \eqref{linMetricClean} looks formally identical to the
  textbook linearisation Einstein equations with an geometric-artificial
  matter term $\delta \Phi_{\mu\nu}$.
  }
\end{IndRemark}

\begin{IndRemark}
  \emph{The initial data for the wave equations \eqref{LinCEFEprop} is
  not free as it needs to satisfy equations \eqref{eqLinearCEFEs} on
  $\Sigma$. The propagation of the (first order) CEFEs by the wave-CEFEs
  has been proven at the non-linear level in \cite{CarHurVal} and the
  result at the linear level follows from linearisation of the
  subsidiary system obtained in \cite{CarHurVal}.  Also, a procedure to
  obtain initial data for the conformal wave CEFEs at the non-linear
  level has been outlined in \cite{CarHurVal}: one starts by solving the
  conformal Einstein constraint equations on $\Sigma$ which constitute
  the spatial parts of the zero-quantities
  \eqref{CFE_tensor_zeroquants}. Then, the time derivatives of the
  evolved fields on $\Sigma$ can be read from the zero-quantities using
  the solution to the conformal Einstein constraints.  An analogous
  procedure can be performed for the linear case to obtain initial data
  for the wave equations \eqref{LinCEFEprop}. Observe that from
  equations \eqref{eq:CEFEconf_factor} and \eqref{eq:CEFE_Riem_Alg_Geo}
  one only needs to consider their trace-free parts as the traces give
  the wave equations \eqref{linCF} and \eqref{linMetricClean},
  respectively. }
\end{IndRemark}

\subsubsection{Scri-fixing gauge}

A difficulty with the standard CEFEs concerns the fact that in general,
the conformal boundary is not fixed with respect to the coordinates.
Although there exist a generalisation of the CEFEs, the \emph{extended
CEFEs} (see for instance \cite{Val16}) that allows to have an explicit
expression the conformal factor, this approach is based on gauge adapted
to a congruence of conformal geodesics, called the conformal Gaussian
gauge. The latter requires the use of more general type of connections
known as Weyl connections and it is in general difficult to write
explicitly the line element of physically relevant solutions in the
coordinates adapted to the conformal Gaussian gauge.  Here, we give an
alternative simpler approach based on exploiting the gauge source
functions in the wave formulation of the standard CEFEs. In the linear
case, the key is to find a gauge choice such that equation \eqref{linCF}
is homogeneous. This amounts to choosing the Lorenz gauge source
functions such that $F^{\mu}$ satisfies
\begin{align}\label{eq:scrifixlinear}
F^\mu \dotnabla_\mu \dotXi = - 4 \delta s + \mathring{s}\delta
g_{\mu}{}^\mu + \frac{1}{6}\dotXi \delta F.
\end{align}
Thus, if trivial initial data is given to equation \eqref{linCF} on the
initial hypersurface then the trivial solution $\delta \Xi=0$ is
obtained in the development. Hence, $\Xi=\dotXi$ and that the location
of the conformal boundary of the perturbed linearised solution coincides
with that of the background spacetime. Observe that to solve equation
\eqref{eq:scrifixlinear} one needs $\dotnabla_\mu \dotXi \neq 0$ and
although this condition does not hold at $i^0$, this is not a problem
for foliations that end at a cut $\mathcal{C}$ of $\mathscr{I}$ (e.g.
the hyperboloidal and null foliations) ---see Figure \ref{fig:diagram}.

In the non-linear case one can proceed in a similar way.  The non-linear
version of the wave equation for the conformal factor reads:
  \begin{align}\label{wave_nl_eq_CF}
    \reducedbox \; \Xi = - \tfrac{1}{6} \Xi F -
    \mathcal{H}^\mu\nabla_\mu \Xi + 4 s
  \end{align}
where $\reducedbox := g^{\mu\nu}\partial_\mu\partial_\nu$ and $x^\mu$ is
a fiduciary coordinate system ---see Appendix \ref{Appendix:B}. One can
follow the same approach as in the linear case and think of $\Xi$ as a
given function of the coordinates and solve for the gauge source
functions. Although this approach is of limited use for the Cauchy
problem as $\nabla_\mu\Xi$ vanishes at $i^0$, this is not an issue for
foliations which end at a cut $\mathcal{C}$ of $\mathscr{I}$ for which
$\nabla_\mu \Xi\neq 0$.  As a concrete example and to make contact with
the hyperboloidal framework of \cite{Zen08, VanHusHil14, Van15, VanHus17,
HilHarBug16} consider an unphysical coordinate system $x^\mu=(\tau, r,
\vartheta^A)$ with $A=1,2$, related to physical coordinates
$\tilde{x}^{\tilde{\mu}} =(\tilde{t}, \tilde{r}, \vartheta^A)$ according
to the standard hyperboloidal prescription $\tilde{t}= \tau +
H(\tilde{r})$ and $\tilde{r}= \frac{r}{\Omega (r)}$ where $H$ and
$\Omega$ are known as the height and compression functions. If one sets
$\Xi= \Omega(r)$ then equation \eqref{wave_nl_eq_CF} fixes one of the
components of the coordinate gauge source functions
  $\mathcal{H}^\mu$ as:
  \begin{align}\label{scri-fixing_coordgauge}
    \mathcal{H}^r = - \frac{\Omega''}{\Omega'}g^{rr} +
    \frac{4s}{\Omega'} - \frac{\Omega}{6\Omega'} F.
  \end{align}
where $\Omega' := \frac{d\Omega}{dr}$. If the slice is hyperboloidal
then $\Omega' \neq 0$ at $\mathscr{I}$.
\begin{IndRemark} \label{rem:NLScrifix}
  \emph{ In \cite{Zen08, Van15, VanHus17} a similar gauge-fixing is
  developed, however in the previous work the evolution equations are
  formally singular while in the CEFE approach of this paper the
  equations are formally regular.  Also, the general strategy put
  forward here is not bound to the hyperboloidal and asymptotically flat
  set-up as a similar scri-fixing procedure could be adapted to study
  spacetimes with de-Sitter (and anti-de-Sitter) asymptotics using CEFEs
  similar to the case of \cite{MinVal23} and \cite{GasVal17} where the
  gauge was fixed using, instead, a congruence of conformal geodesics.}
\end{IndRemark}
%
\begin{IndRemark}
  \emph{ Recall that $g_{ab}=\Xi^2\tilde{g}_{ab}$, hence for the linear
  perturbations one has:
  \begin{align}\label{Rel_Metric_Perts2}
    \delta g_{\mu\nu}=\mathring{\Xi}^2\delta \tilde{g}_{\mu\nu} +
    2\mathring{\Xi}\mathring{\tilde{g}}_{\mu\nu}\delta \Xi .
  \end{align}
  Using the above described gauge to set $\delta \Xi=0$, then equation
  \eqref{Rel_Metric_Perts2} reduces to $ \delta
  g_{\mu\nu}=\mathring{\Xi}^2 \delta \tilde{g}_{\mu\nu}$.  However, it
  should be stressed that $\delta \tilde{g}_{\mu\nu} \neq \delta
  \tilde{g}_{\tilde{\mu}\tilde{\nu}}$. Namely, $\delta
  \tilde{g}_{\mu\nu}$ denotes the components of the physical metric
  perturbation $\delta \tilde{g}_{ab}$ in the unphysical Cartesian
  coordinate basis $\bm\partial_\mu$. Using equation
  \eqref{CompsRelation} one sees that the expression for the physical
  metric perturbation in the physical Cartesian coordinate basis
  $\bm\partial_{\tilde{\mu}}$ is given by
  \begin{align}
    \delta \tilde{g}_{\tilde{\mu}\tilde{\nu}} =
    (\delta_{\tilde{\mu}}^\mu -2x^\mu\tilde{x}_{\tilde{\mu}})
    (\delta_{\tilde{\nu}}^\nu -2x^\nu\tilde{x}_{\tilde{\nu}}) \delta
    g_{\mu \nu} .
  \end{align}
  }
\end{IndRemark}

\subsection{General solutions to the linearised equations}
\label{sec:gen_sol_lin}

\subsubsection{The vacuum equations}
In this subsection an analysis of the linearisation of the wave CEFEs in
vacuum is given (later, we extend the discussion to the case where
matter is included and with a more general gauge choice). To start the
discussion in the simplest possible set up, consider the vacuum case
with vanishing gauge source functions $F_{\mu}=\delta F=0$ so that the
equations read
\begin{subequations}\label{FLCEFE-LCEFEs1}
    \begin{align}
      \mathring{\square}\delta  g_{\{\mu  \nu\}  } & = 
      -4 \delta \Phi _{\mu  \nu  }, \label{FLCEFEmetricsplit1}\\
      \mathring{\square} \delta \Xi  & = 4 \delta s  -
      \mathring{s} \delta g^{\mu  }{}_{\mu  },  \label{FLCEFEXi}\\
      \mathring{\square}\delta g^{\mu  }{}_{\mu  } & = 0 ,
      \label{FLCEFEmetricsplit2} \\
      \mathring{\square}\delta s & =0, \label{FLCEFE-LCEFEh1}\\
      \mathring{\square}\delta \Phi_{\mu  \nu  } & =0 ,
      \label{FLCEFE-LCEFEh2} \\
      \mathring{\square}\delta d_{\mu  \nu \alpha \beta  } & =0  
      \label{FLCEFE-LCEFEh3},
    \end{align}
\end{subequations}
subject to the constraints imposed by equations \eqref{eqLinearCEFEs}
with vanishing matter terms. Note that we have split the metric
perturbation into its trace $\delta g_\mu{}^\mu$ and its trace-free part
$\delta g_{\{\mu\nu\}}$. In this manner, one can identify the variables
that satisfy homogeneous equations.

\subsubsection{Solutions to the homogeneous equations}

The solutions of homogeneous wave equations in flat spacetime are widely
known and understood, but since the initial data are subject to the
initial constraints \eqref{eqLinearCEFEs}, it is perhaps appropriate to
briefly discuss how one might write the solutions explicitly in terms of
the initial data. For a field $\delta{\bmQ}_H$ satisfying a homogeneous
wave equation $\mathring{\square}\delta{\bmQ}_H=0$, the solution may be
written in the form \cite{BarutElectrodynamics}:
\begin{align}\label{FLCEFE-Soln}
  \delta {\bmQ}_H(x) 
  &= - \int_\Sigma d \Sigma^\prime_\lambda 
      \left[ \underline{\delta \bmQ}_H(x^\prime) 
        \partial^\lambda D(x,x^\prime)
        - D(x,x^\prime) 
        \underline{\partial^\lambda \delta \bmQ}_H(x^\prime)
      \right] ,
\end{align}
where $d \Sigma^\prime_\lambda = (1/3!)\epsilon_{\lambda \alpha \beta
\gamma} dx^{\prime\alpha} \wedge dx^{\prime\beta} \wedge
dx^{\prime\gamma}$ is the surface element on $\Sigma$, the underline
denotes evaluation of a quantity on $\Sigma$, and the quantity
$D(x,x^\prime):=G^+(x,x^\prime)-G^-(x,x^\prime)$ is the difference
between the retarded ($-$) and advanced ($+$) Green's functions, defined
as 
\begin{align} \label{AdvRetGreen}
  G^{\pm}(x,x^\prime)
  &:= - \frac{\delta(x^{0}-x^{\prime 0}\pm|\vec{x}-\vec{x}^\prime|)}
          {4\pi |\vec{x}-\vec{x}^\prime|}
  ,
\end{align}
and satisfying the inhomogeneous equation $\mathring{\square}
G^{\pm}(x,x^\prime) = -\delta^4(x-x^\prime)$. That \eqref{FLCEFE-Soln}
is a solution of (\ref{FLCEFE-LCEFEh1}-\ref{FLCEFE-LCEFEh3}) can be seen
by noting that $D(x,x^\prime)$, being the difference of two Green's
functions, is a solution of the homogeneous wave equation. It is
straightforward to show that $D(x,x^\prime)=0$ if $x^{0}=x^{0\prime}$,
and one can also show \cite{BarutElectrodynamics} that $D(x,x^\prime)$
satisfies the properties:
\begin{subequations}\label{GreenDiffProp}
\begin{align}
  -\left.
  \int_\Sigma d \Sigma^\prime_\tau 
        \left[ 
          \underline{\bmQ}_H(x^\prime) 
          \partial^\tau D(x,x^\prime) 
    \right]
  \right|_{x \in \Sigma} &= \underline{\bmQ}_H(x \in \Sigma) ,
  \label{GreenDiffPropa}\\
  \left.
            \frac{\partial^2 D(x,x^\prime)}{\partial \Delta t^2}  
  \right|_{\Delta t=0} &= 0
  \label{GreenDiffPropb}
  .
\end{align}  
\end{subequations} 
where $\Delta t:=x^{0}-x^{\prime 0}$. Equation \eqref{GreenDiffPropb}
follows from the fact that for a given $|\vec{x}-\vec{x}^\prime|$,
$D(x,x^\prime)$ is by construction an odd function in $\Delta t$. Given
\eqref{GreenDiffProp}, one can verify that the coefficients in
\eqref{FLCEFE-Soln} do in fact coincide with the initial data (though
note that \eqref{FLCEFE-Soln} only depends on the time derivatives of
the fields). Of course, as indicated earlier, the initial data for the
fields must satisfy the vacuum version of \eqref{eqLinearCEFEs}.

Given the simplicity of the vacuum wave equations and their resemblance
to wave equations in ordinary Minkowski space it is tempting to write
the solutions in terms of plane waves. While one can certainly obtain
solutions to the wave equations in this manner, a difficulty arises from
the fact that even in the vacuum case, the constraints
\eqref{eqLinearCEFEs} are nontrivial, since the background expression
for $\mathring{\Xi}$ in the inversion coordinates is explicitly
coordinate dependent. As a result, solutions satisfying the constraints
do not have a clean mode separation in a Cartesian Fourier expansion
(though perhaps an alternative mode decomposition may be appropriate).

\subsubsection{Solutions to the inhomogeneous equations and including 
matter}

For completeness, we discuss the solutions of equations
\eqref{FLCEFEmetricsplit1} and \eqref{FLCEFEXi}. One may begin by
assuming that one has in hand initial data satisfying the constraints
\eqref{eqLinearCEFEs}. We note that the structure of the vacuum
equations has a hierarchy: one can first solve for $\delta s$, $\delta
\Phi_{\mu \nu}$, $\delta d_{\mu \nu \alpha \beta}$, and $\delta
g^{\mu}{}_{\mu}$, then use the result to solve for $\delta g_{\{\mu
\nu\}}$ and $\delta \Xi$. Schematically, one may then write the
inhomogeneous wave equations for the latter two in the following form
(where ${\bm S}$ is a source that depends on the background field
expressions and $\delta s$, $\delta \Phi_{\mu \nu}$, and $\delta
g^{\mu}{}_{\mu}$):
\begin{equation}\label{FLCEFE-IHwave}
    \mathring{\square}\delta {\bmQ}_I(x) = {\bm S}(x),
\end{equation}
which yields a particular solution (with $x^0=t$, 
$x^{\prime 0}=t^\prime$):
\begin{equation}\label{FLCEFE-SolnIH}
  \begin{aligned}
      \delta {\bmQ}_{I_0}(x) 
      &= \delta {\bmQ}_{H_0}(x) 
        - \int_{t_0}^t dt^\prime \int d^3\vec{x}^\prime 
        \left[ 
          G^{-}(x,x^\prime) {\bm S}(x^\prime)
        \right] ,
  \end{aligned}
\end{equation}
where $\delta {\bmQ}_{H_0}(x)$ is a solution to the homogeneous equation
with initial conditions chosen so that the particular solution $\delta
{\bmQ}_{I_0}(x)$ has trivial initial conditions $\delta
{\bmQ}_{I_0}(x)=0$ and $\partial_t \delta {\bmQ}_{I_0}(x)=0$ on
$\Sigma$; this can in principle be done by evaluating the integral term
and its derivative on $\Sigma$ to obtain $-\underline{\delta
\bmQ}_{H_0}(x^\prime)$ and $-\underline{\partial_t \delta
\bmQ}_{H_0}(x^\prime)$. Finally, to obtain the solution corresponding to
the prescribed initial data, one can then add to $\delta
{\bmQ}_{I_0}(x)$ a homogeneous solution of the form in equation
\eqref{FLCEFE-Soln}. Since we are interested in the initial value
problem, we have chosen the retarded Green's function
$G^{-}(x,x^\prime)$ in \eqref{FLCEFE-SolnIH}. This is (as the reader
might be aware) because $G^{-}(x,x^\prime)$ is only nontrivial for $x^0
> x^{\prime 0}$, so that for a given point $x$, the field contributions
from the integral in \eqref{FLCEFE-SolnIH} depend on the values of the
source in the past of $x$ (specifically over the past light cone of
$x$).

We note that the procedure described for the inhomogeneous equation can
be straightforwardly extended to the full set of non-vacuum equations
\eqref{LinCEFEprop}, subject to the constraints \eqref{eqLinearCEFEs} on
the initial data. As in the vacuum case, one has a hierarchy in which
one can first specify the matter sources, then solve for the variables
$\delta s$, $\delta \Phi_{\mu \nu}$, $\delta d_{\mu \nu \alpha \beta}$,
and $\delta g_{\mu \nu}$, and finally solve for the conformal
perturbation $\delta \Xi$ using the previously obtained results. Of
course, the preceding discussion glosses over the problem of solving the
matter equations, however such analysis is matter-model dependent. Hence
in the present analysis they are simply regarded as given sources in the
equations.

\subsection{The far-field approximation and quadrupole-like formulae}
\label{sec:quadrupolelike}

As emphasised before, an appealing property of the wave formulation of
the CEFEs is that the metric sector of the equations looks as the
standard non-conformal Einstein field equations with some
artificial-geometric matter term. This suggests that some of the
standard techniques and approximation methods can be applied to the
linearised wave CEFEs such as solving them in the far-field regime and
obtaining quadrupole-like formulae.

\subsubsection{The far-field approximation}

To derive the far-field approximation, we focus our attention not to the
homogeneous part of the solution to equation \eqref{FLCEFE-IHwave} but
to the part generated by the sources. For simplicity, we omit nontrivial
homogeneous solutions and $\delta {\bmQ}_{H_0}(x)$ in the particular
solution (which corresponds to an appropriate choice of initial data),
so that the solution consists only of the integral in equation
\eqref{FLCEFE-SolnIH}. Upon evaluating the time integral, one obtains:
\begin{align}\label{eq:gen_sol_wave}
   \delta \bmQ= -\frac{1}{4\pi}\int_{\Sigma}
   \frac{\bmS(t-|\vec{x}-\vec{x}^\prime|,\vec{x}^\prime)}
   {|\vec{x}-\vec{x}^\prime|}d^3\vec{x}^\prime.
\end{align}
This is of course just the standard expression one can find in standard
textbooks---see  \cite{Mag07a}.  To obtain a far-field expression in the
unphysical set-up we start by recalling that the origin in the inversion
(unphysical) Minkowski spacetime corresponds to spatial infinity $i^0$
for the physical Minkowski spacetime so the far-field region in physical
Minkowski spacetime corresponds to a neighbourhood of the origin in the
inversion Minkowski spacetime.  Let $\bmp'$ and $\bmp$ be a points in
the physical Minkowski spacetime with physical Cartesian coordinates
$(0, \tilde{x}^{\prime i})$ and $(0, \tilde{x}^i)$ respectively.
Expressed in unphysical Cartesian coordinates, these points correspond
to $(0, x^{\prime i})$ and $(0,x^i)$. Hence, the distance $r$ between
$\bmp'$ and $\bmp$ can be expressed through the vector

 \begin{align}
   \vec{r}:=\vec{x}-\vec{x}^\prime=\vec{\tilde{x}}
      -\vec{\tilde{x}}^{\prime}.
 \end{align}
In terms of the physical coordinates, imposing the far-field
approximation means $|\vec{\tilde{x}}|>>|\vec{\tilde{x}}^\prime|$.
Consequently,
 \begin{align}
   r:=|\vec{r}| = |\vec{\tilde{x}}-\vec{\tilde{x}}^\prime| 
   \simeq |\vec{\tilde{x}}|
   =\frac{1}{|\vec{x}|}= \frac{1}{\rho}.
 \end{align}
Therefore, in the \emph{far-field approximation} one can write
  \begin{align}\label{gen_far_field}
    \delta \bmQ \simeq -\frac{\rho}{4\pi}\int_{\mathbb{R}^3} \bmS 
    (t-r,\vec{x}^\prime)d^3\vec{x}^\prime.
  \end{align}
In particular, for the case where $\delta \bmQ$ is either $\delta
\Phi_{\mu\nu}$, $\delta g_{\mu\nu}$ or $\delta d_{\alpha\beta\mu\nu}$
one can derive quadrupole-like formulae. To do so, it is necessary to
compute the divergence of the source terms appearing in equations
\eqref{linMetricClean}, \eqref{linTFRicci} and \eqref{linResWeyl}.
Let's define
\begin{align}\label{eq:sourcesdefs}
    \mathcal{S}_{\mu\nu} & :=-4 \delta \Phi _{\mu  \nu  }
    - \mathring{g}_{\mu  \nu  } \dotnabla _{\alpha  }F^{\alpha  }
    + \dotnabla _{\mu  }F_{\nu  } + \dotnabla _{\nu  }F_{\mu  } +
    \tfrac{1}{2}\delta F \dotg_{\mu\nu} \\
   S_{\mu\nu} &:=-2 \delta T_{(\mu}{}^\alpha{}_{\nu)}\dotnabla_\alpha \dotXi 
  - \dotXi \dotnabla_\alpha \delta T_{\mu}{}^\alpha{}_{\nu}
  + \frac{1}{6}\dotnabla_{\{\mu}\dotnabla_{\nu\}} \delta F, \\
   S_{\mu\nu\alpha\beta} &:=
  2 \dotnabla_{[\mu} \delta T _{|\alpha\beta|\nu]} 
  + 2 \dotnabla_{[\alpha} \delta T_{|\mu\nu|\beta]} 
  + 2 \dotg_{\beta[\nu}\dotnabla_{|\sigma}
      \delta T_{\alpha|}{}^{\sigma}{}_{\mu]} 
  + 2 \dotg_{\alpha[\mu}\dotnabla_{|\sigma}
      \delta T_{\beta|}{}^{\sigma}{}_{\nu]}.
\end{align}
Recall that the fields encoded in $\mathring{\bm\phi}$ satisfy the
CEFEs, hence the background conformal factor satisfies
$\mathring{Z}_{ab}=0$ as defined in equation
\eqref{StandardCEFEsecondderivativeCF}, which for the flat background
case reduces to $\dotnabla _{\mu}\dotnabla _{\nu}\dotXi =  \mathring{s}
\dotg_{\mu\nu}$. A direct calculation using the latter expression  and
the matter relations \eqref{eq:LinMatterGeneralRels} shows that
\begin{align}\label{eq:DivSs}
  \dotnabla^\nu \mathcal{S}_{\mu\nu}=\mathring{\square}F_{\mu}, \qquad
  \dotnabla^\nu  S_{\mu\nu}= 
  \tfrac{1}{8}\mathring{\square}\dotnabla_\nu \delta F, \qquad
  \dotnabla^\lambda  S_{\lambda\alpha\sigma\beta} =
  \mathring{\square} \delta T_{\sigma\beta\alpha}.
\end{align}
The first two equations simply encode equations
\eqref{eq:lorenzpropmain} and \eqref{eq:linearisedConfConsProp}  in a
divergence-like format. The third equation in \eqref{eq:DivSs} shows
that equation \eqref{eq:CEFElinWeyl} propagates.

\begin{IndRemark}
  \emph{
  For the scalar variables $\delta s$ and $\delta \Xi$, the leading
  order source terms can be evaluated directly. In the case of rank-2
  tensors, a reason for reducing the sources to quadrupole form is
  to make the field dependent on a single (the time-time) component of 
  the source tensor.
  }
\end{IndRemark}

\subsubsection{A quadrupole-like formula for $\delta \Phi_{\mu\nu}$ and $\delta g_{\mu\nu}$}
\label{sec:quad_met}

Here, we derive quadrupole-like formulae for $\delta \Phi_{\mu\nu}$ and
$\delta g_{\mu\nu}$, which we discuss in tandem. Since the source term
$\mathcal{S}_{\mu\nu}$ contains $\delta \Phi_{\mu\nu}$ [cf. equation
\eqref{eq:sourcesdefs}] first one solves for $\delta \Phi_{\mu\nu}$
and then for $\delta g_{\mu\nu}$.  Let $ \delta Q_{\mu\nu} = \delta
\Phi_{\mu\nu} \; \text{or} \;\delta g_{\mu\nu}$ and $\mathcal{F}_{\mu} =
\tfrac{1}{8}\dotnabla_\mu \delta F \; \text{or}\; F_{\mu}$, and
$\sigma_{\mu\nu} = S_{\mu\nu} \; \text{or} \; \mathcal{S}_{\mu\nu}$,
respectively.  Exploiting that equations \eqref{LinBianchiContracted}
and \eqref{generalisedLorenzGauge} can be collectively written as
$\dotnabla^\nu \delta Q_{\mu\nu}= \mathcal{F}_{\mu}$ one has that
\begin{align}\label{eq:gendivexpanded}
  \dotnabla_0 \delta Q_{00}=\dotnabla^i\delta Q_{i0} -\mathcal{F}_{0},
  \qquad \dotnabla_0 \delta Q_{0i}=\dotnabla^j\delta Q_{ij}
  -\mathcal{F}_{i},
\end{align}
Thus, once $\delta Q_{ij}$ is determined, $\delta Q_{0i}$ can be
obtained by formally integrating the second equation in
\eqref{eq:gendivexpanded}. In turn, $\delta Q_{00}$ is obtained by
integrating the first equation in \eqref{eq:gendivexpanded}.  For the
particular case of $\delta \Phi_{\mu\nu}$ one could instead use that
$\delta \Phi_\mu{}^\mu=0$ to solve algebraically for $\delta \Phi_{00}$.
Thus, one can reduce the discussion to that of obtaining a
quadrupole-like formula for $\delta Q_{ij}$. Recall that the far-field
approximation reads
\begin{align}\label{eq:farfieldExpg}
  \delta Q_{ij} \simeq -\frac{\rho}{4\pi}\int_{\Sigma}
  \sigma_{ij}(t-r,\vec{x}^\prime)d^3\vec{x}^\prime.
\end{align}
If one were to restrict the calculations to gauge source functions
satisfying $\mathring{\square}F^\mu = \mathring{\square} \dotnabla_\mu
\delta F= 0$ (in particular, $F^\mu= \delta F=0$) then the integrand
would satisfy $\dotnabla^\nu \mathcal{\sigma}_{\mu\nu}=0$ leading to an
expression which is formally identical to the quadrupole formula in the
non-conformal set-up for the physical metric perturbation ---see for
instance \cite{Mag07a, Wal84}.  However, it is instructive to keep the
gauge source function terms since it will serve as model for the
derivation of the quadrupole-like formula for the rescaled Weyl tensor
where the divergence of the source term $S_{\mu\nu\alpha\beta}$ cannot
be set to zero by gauge considerations ---see the last equation in
 \eqref{eq:DivSs}. Here, we only assume that the gauge source
functions are specified. The first two equations \eqref{eq:DivSs} in the
current notation read $\dotnabla^\nu \sigma_{\mu\nu} =
\mathring{\square} \mathcal{F}_{\mu} $, which in components read
\begin{align}\label{eq:divSigma}
  \dotnabla_{0} \sigma^{00} + \dotnabla_{i} \sigma^{i0}
  =\mathring{\square} \mathcal{F}^0, \qquad \dotnabla_{0} \sigma^{0i}
  + \dotnabla_{j} \sigma^{ij} = \mathring{\square} \mathcal{F}^i.
\end{align}
Integrating $\dotnabla_i(\sigma^{ik}x^{\prime j})$ over a region
$\Omega \subset \Sigma$ and using Gauss' law gives the following
identity
\begin{align}\label{eq:boundaryDeltaSigma1}
  \int_{\partial\Omega} \sigma^{ik} x^{\prime j} ds_i =\int_\Omega
  \sigma^{jk}d^3\vec{x}^\prime + \int_{\Omega} x^{\prime j}\dotnabla_i
  \sigma^{ik}d^3\vec{x}^\prime,
\end{align}
where we have used $\dotnabla_i x^{\prime j} = \delta{_i}{^j}$ and
$ds_i$ is the surface element of $\partial\Omega$.
\begin{IndAssumption}\label{matter-support1}
  \emph{ It will be assumed that the matter fields encoded in
  $S_{\mu\nu}$ and $S_{\mu\nu\alpha\beta}$, as defined in equation
  \eqref{eq:sourcesdefs}, have compact support in $\Omega \subset
  \mathbb{R}^3$.  }
\end{IndAssumption}

\begin{IndRemark}\label{compactsuportphi}
  \emph{
Recall that  $\mathring{\square}\Phi_{\mu\nu}=S_{\mu\nu}$, then
under assumption \ref{matter-support1} it follows that
$\delta \Phi_{\mu\nu}$ has compact support in $\Omega$.
Furthermore, choosing
the gauge source function $F^\mu$ appropriately,
one can then assume that that $\mathcal{S}_{\mu\nu}$ [as given in equation
\eqref{eq:sourcesdefs}] has compact support in $\Omega$.
}
\end{IndRemark}

Using assumption \ref{matter-support1} and Remark
\ref{compactsuportphi}, equations \eqref{eq:boundaryDeltaSigma1} and
\eqref{eq:divSigma} give
\begin{align}\label{eq:intquadsmet1}
  \int_{\Omega}
  \sigma^{jk}d^3\vec{x}^\prime=-\int_{\Omega}x^{\prime(j}
  \mathring{\square}\mathcal{F}^{k)}d^3\vec{x}^\prime +
  \frac{d}{dt^{\prime }}\int_{\Omega}x^{\prime(j}
  \sigma^{k)0}d^3\vec{x}^\prime.
\end{align}
Similarly, integrating $\dotnabla_k(\sigma^{0k}x^{\prime i}x^{\prime
  j})$ over a region $\Omega$ and proceeding as before, making use of
Gauss' law one obtains
\begin{align}\label{eq:boundaryDeltaSigma2}
  \int_{\partial\Omega} \sigma^{0k}x^{\prime i}x^{\prime j} ds_k
  =\int_\Omega x^{\prime i}x^{\prime j} \dotnabla_k
  \sigma^{0k}d^3\vec{x}^{\prime} + 2 \int_{\Omega} x^{\prime (i}
  \sigma^{j)0}d^3\vec{x}^\prime.
\end{align}
Then using Remark \ref{compactsuportphi} and equations
\eqref{eq:boundaryDeltaSigma2} and \eqref{eq:divSigma} renders
\begin{align}\label{eq:intquadsmet2}
  \int_{\Omega} x^{\prime(i} \sigma^{j)0}d^3\vec{x}^\prime  =
  -\frac{1}{2} \int_\Omega x^{\prime i}x^{\prime j}
  \mathring{\square}\mathcal{F}^0 d^3\vec{x}^\prime
  + \frac{1}{2}\frac{d}{dt^{\prime}}
  \int_\Omega x^{\prime i}x^{\prime j} \sigma^{00}d^3\vec{x}^\prime
\end{align}
Combining equations \eqref{eq:intquadsmet1}
and \eqref{eq:intquadsmet2} one gets
\begin{align}
   \int_{\Omega}  \sigma^{jk}d^3\vec{x}^\prime =
     \frac{1}{2}\frac{d^2}{d{t^{\prime}}^2}
   \int_\Omega x^{\prime i} x^{\prime j} 
   \sigma^{00}d^3\vec{x}^\prime-\frac{1}{2} \frac{d}{dt^\prime}
   \int_\Omega x^{\prime i}x^{\prime j} 
   \mathring{\square}\mathcal{F}^0 d^3\vec{x}^\prime
   -\int_{\Omega}x^{\prime(j}
   \mathring{\square}\mathcal{F}^{k)}d^3\vec{x}^\prime
\end{align}
therefore, using the far-field approximation \eqref{eq:farfieldExpg}
gives
\begin{align}\label{eq:PhiQuadrupole2}
  \delta Q_{ij} \simeq -\frac{\rho }{8\pi}
  \frac{d^2\mathcal{I}_{ij}(t')}{dt'^2}\Bigg|_{t'=t-r} +\frac{\rho
  }{8\pi} \frac{d\mathcal{J}_{ij}(t')}{dt'}\Bigg|_{t'=t-r} -\frac{\rho
  }{4\pi} \mathcal{K}_{ij}(t')\Bigg|_{t'=t-r}
\end{align}
where
\begin{align}\label{eq:defQuadrupoleSigma}
  \mathcal{I}^{ij}:= \int_{\Omega} \sigma^{00}x^{\prime i}x^{\prime
    j}d^3\vec{x}^\prime, \qquad \mathcal{J}^{ij} := \int_{\Omega}
  x^{\prime i}x^{\prime j}
  \mathring{\square}\mathcal{F}^{0}d^3\vec{x}^\prime, \qquad
  \mathcal{K}^{ij} := \int_{\Omega}
  \mathring{\square}\mathcal{F}^{(i}x^{\prime j)} d^3\vec{x}^\prime.
\end{align}

As discussed earlier, the gauge source functions could be set to zero, 
but in certain applications, it can be useful to consider general gauge 
source functions. Of course, one may wish to implement a scri-fixing 
gauge for this situation, but such a gauge may be incompatible with the
assumption that $\mathcal{S}_{\mu\nu}$ has compact support in $\Omega$.
However, the purpose of the present analysis is to provide a template
for deriving the quadrupole formula for the Weyl tensor, in which such a
scri-fixing gauge can be implemented.

\begin{IndRemark}
  \emph{
  The reader might wonder whether quadrupole-like formulae are
  appropriate for trace-free energy-momentum tensors, which often
  corresponds to that of radiation. The concern is that in order to
  implement the far-field approximation, one implicitly assumes a
  slow-motion approximation, which is inconsistent with radiation
  propagating at the speed of light. However, geons
  \cite{Wheeler1955geons,BrillHartle1964geon,AndersonBrill1997geon} and
  photon stars
  \cite{Sorkin:1981wd,Schmidt:1999tr,mitra2010likely,Kim:2016jfh}
  provide examples of stationary compact sources formed from
  self-gravitating radiation, which may at the very least provide useful
  toy models (the extension of our results to the case of nonvanishing
  trace will be discussed elsewhere). Nevertheless, the primary reason
  for the preceding analysis is to provide a template for deriving the
  quadrupole formula for the rescaled Weyl tensor, which can be
  straightforwardly generalised to the case with trace. This is discussed
  further in Remark \ref{Rem:GenTrace-Free}. 
  }
\end{IndRemark}

\subsubsection{A quadrupole-like formula for the rescaled Weyl tensor 
perturbation}
\label{sec:quad_Weyl}
We now turn to the case of the rescaled Weyl tensor $\delta
Q_{\alpha\beta\mu\nu}=\delta d_{\alpha\beta\mu\nu}$. Here, the source
term $S_{\alpha\beta\mu\nu}$ is given entirely in terms of matter fields
---see equation \eqref{eq:sourcesdefs}. Expanding in components the
divergence equation \eqref{eq:DivSs} one has
\begin{align}\label{eq:DivSTspatial}
  \dotnabla_0 S^{0\mu\sigma\nu} + \dotnabla_k
  S^{k\mu\sigma\nu} = \mathring{\square} \delta
  T^{\sigma\nu\mu}
\end{align}
Considering the case $\sigma=j$ in equation \eqref{eq:DivSs} one has
\begin{align}\label{eq:DivSTspatial1}
  \dotnabla_0 S^{0\mu j \nu} + \dotnabla_k S^{k\mu j \nu} =
  \mathring{\square} \delta T^{j \nu\mu}
\end{align}
Integrating $\dotnabla_k ( S^{k \mu j \nu} x^{\prime i})$ over $\Omega$
and using Gauss' law gives
\begin{align}\label{eq:weylquadrupole1}
\int_{\partial\Omega} S^{k \mu j \nu}x^{\prime i} ds_k = 
  \int_\Omega S^{i\mu j\nu} d^3\vec{x}^\prime 
  + \int_\Omega x^{\prime i} \dotnabla_k S^{k \mu j \nu}
  d^3\vec{x}^\prime
\end{align}
Using assumption \ref{matter-support1} the boundary term in the
left-hand side of \eqref{eq:weylquadrupole1} vanishes. Solving for the
integral with $S^{i\mu j \nu}$ and substituting $\dotnabla_k S^{k \mu
  j \nu}$ using equation \eqref{eq:DivSTspatial1}, gives the following
expression (after symmetrising the spatial indices):
\begin{align}\label{eq:weylquadpiece1}
  \int_\Omega S^{(i |\mu| j) \nu}d^3\vec{x}^\prime =
  - \int_\Omega x^{\prime (i} \mathring{\square} \delta T^{j)\nu\mu}
  d^3\vec{x}^\prime
  +  \frac{d}{dt^\prime}\int_{\Omega}  x^{\prime (i|}  S^{0\mu |j) \nu}
  d^3\vec{x}^\prime .
\end{align}
Now, considering the case $\sigma =0$ in equation \eqref{eq:DivSs} one
has
\begin{align}\label{eq:DivSTspatial2}
  \dotnabla_0  S^{0\nu 0 \mu}  + \dotnabla_k  S^{k \nu 0 \mu}
  =  \mathring{\square} \delta T^{0 \mu\nu}.
\end{align}
Integrating  $\dotnabla_k ( S^{0 \mu k \nu} x^{\prime i}x^{\prime j})$
over $\Omega$ and using Gauss' law gives
\begin{align}\label{eq:weylquadrupole2}
  \int_{\partial\Omega} S^{0 \mu k \nu}x^{\prime i}x^{\prime j} ds_k =
  2 \int_\Omega  S^{0 \mu (j |\nu|} x^{\prime i)} d^3\vec{x}^\prime
  + \int_\Omega x^{\prime i}x^{\prime j} 
  \dotnabla_k  S^{0 \mu k \nu} d^3\vec{x}^\prime.
\end{align}
Using assumption \ref{matter-support1} to remove the boundary terms and
exploiting the pair-interchange-symmetry of $S_{\mu\nu\alpha\beta}$ to
rewrite $\dotnabla_k S^{0 \mu k \nu}$ in using equation
\eqref{eq:DivSTspatial2} renders
\begin{align}\label{eq:weylquadpiece2}
  \int_\Omega S^{0 \mu (j |\nu|} x^{\prime i)}d^3\vec{x}^\prime =
  -\frac{1}{2} \int_\Omega x^{\prime i}x^{\prime j}
  \mathring{\square}\delta T^{0 \mu\nu} d^3\vec{x}^\prime +
  \frac{1}{2} \frac{d}{dt^\prime}\int_\Omega x^{\prime i}x^{\prime j}
  S^{0\nu 0 \mu}d^3\vec{x}^\prime.
\end{align}
Together, equations \eqref{eq:weylquadpiece1} and
\eqref{eq:weylquadpiece2} give
\begin{align}\label{eq:weylquadmain}
  \int_\Omega S^{(i |\mu| j) \nu} d^3\vec{x}^\prime = \frac{1}{2}
  \frac{d^2}{d{t^\prime}^2}\int_\Omega x^{\prime i}x^{\prime j}
  S^{0\nu 0 \mu}d^3\vec{x}^\prime -\frac{1}{2}\frac{d}{dt^\prime}
  \int_\Omega x^{\prime i}x^{\prime j} \mathring{\square} \delta T^{0
    \mu\nu}d^3\vec{x}^\prime - \int_\Omega x^{\prime (i}
  \mathring{\square}\delta T^{j)\nu\mu} d^3\vec{x}^\prime.
\end{align}
To employ equation \eqref{eq:weylquadmain} one needs to perform an
electric-magnetic decomposition of the tensors involved in the current
discussion.  Consider any tensor $W_{abcd}$ with the symmetries of the
Weyl tensor and let $n^a$ be a normalised timelike vector such that
$n_an^a=-1$. Define the associated projector as
$h_{a}{}^{b}:=g_{a}{}^{b} + n_an^b$. Then, the electric part $E_{ab}$
and magnetic part $B_{ab}$ of $W_{abcd}$ are given by
\begin{align}
E_{ab}:=W_{efgd}n^fn^dh_{a}{}^eh_{b}{}^d, \qquad
B_{ab}:=W^*_{efgd}n^fn^dh_{a}{}^eh_{b}{}^d,
\end{align}
where $B_{ab}=-\tfrac{1}{2}\epsilon_{cd}{}^{ef}W_{abef}$ with
$\epsilon_{abcd}$ is the (4-dimensional) volume-form of $g_{ab}$.  In
practice (particularly when working in an adapted frame), instead of
working directly with $B_{ab}$ it is simpler to use
\begin{align}
  B_{abc}:=W_{efgh}n^fh_{a}{}^eh_{b}{}^gh_{c}{}^h,
\end{align}
which is related to $B_{ab}$ via $B_{ab}= -
\frac{1}{2}B_{acd}\epsilon_{b}{}^{cd} $ where
$\epsilon_{abc}:=\epsilon_{fabc}n^f$ is the (3-dimensional) volume-form
of $h_{ab}$ ---see \cite{Val16, Fri96} for further discussion on the
electric-magnetic decomposition. In the present case, since the
background is flat, one can simply take the normal vector to be $n^\nu=
\delta_0{}^\nu$ so that the electric-magnetic split of
$d_{\mu\nu\alpha\beta}$ simply correspond to
\begin{align}
\delta d_{ij}:= \delta d_{i0j0}, \qquad \delta d_{ijk}:=\delta
d_{i0jk}.
\end{align}
The wave equation for $\delta d_{\mu\nu\alpha\beta}$, which compactly
written  reads $\mathring{\square}\delta d_{\mu\nu\alpha\beta}=
S_{\mu\nu\alpha\beta}$, can then be split into the wave equations for
$d_{ij}$ and $d_{ijk}$:
\begin{align}
\mathring{\square}\delta d_{ij}= S_{ij}, \qquad
\mathring{\square}\delta d_{ijk}= S_{ijk},
\end{align}
where $ S_{ij}:=  S_{i0j0} $ and $ S_{ijk}:= S_{i0jk}$. Using the
far-field approximation expression \eqref{gen_far_field} and equation
\eqref{eq:weylquadmain} then gives
\begin{align}
 \delta d_{i\mu j \nu } \simeq -\frac{\rho}{8\pi}
 \frac{d^2I_{ij\mu\nu}(t')}{dt'^2}\Bigg|_{t'=t-r} +\frac{\rho}{8\pi}
 \frac{dJ_{ij\mu\nu}(t')}{dt'}\Bigg|_{t'=t-r} +\frac{\rho}{4\pi}
 K_{ij\mu\nu}(t')\Bigg|_{t'=t-r},
\end{align}
where
\begin{align}
  I^{ij \mu\nu}:=\int_\Omega y^iy^j \delta S^{0\nu 0 \mu}d^3\vec{y},
  \qquad J^{ij \mu\nu}:=\int_\Omega y^iy^j \mathring{\square} \delta
  T^{0 \mu\nu}d^3\vec{y}, \qquad K^{ij \mu\nu}:= \int_\Omega y^{(i}
  \mathring{\square} \delta T^{j)\nu\mu} d^3\vec{y}.
\end{align}
In particular, setting $\mu=0$ and $\nu=0$ one gets the quadrupole-like
expression for $\delta d_{ij}$ and similarly taking $\mu=0$ and $\nu =
k$ for $\delta d_{ijk}$.  Proceeding as described before and exploiting
the symmetries of $S^{\mu\nu\alpha\beta}$ (those of the Weyl tensor) and
those of $\delta T^{\alpha\mu\nu}$ [cf. equation
\eqref{eq:LinMatterGeneralRels}] renders
\begin{align}\label{eq:quadElectricMagnetic}
  \delta d_{ij}   \simeq \frac{\rho}{4\pi} K_{ij00}(t')\Bigg|_{t'=t-r} ,
  \qquad
  \delta d_{ijk}   \simeq \frac{\rho}{4\pi} K_{ijk0}(t')\Bigg|_{t'=t-r},
\end{align}
Recall that the rescaled Cotton tensor $T_{\alpha \mu\nu}$ can be
expressed in terms of the derivatives of the unphysical energy momentum
tensor. Therefore equation \eqref{eq:quadElectricMagnetic} allows one to
express the perturbation to the rescaled Weyl tensor in the far-field
approximation in terms  derivatives of the perturbation to the
unphysical energy momentum tensor $\delta T_{\mu\nu}$ and background
quantities via equation \eqref{eq:CottToEne}.

\begin{IndRemark} \label{Rem:GenTrace-Free}
  \emph{ Although the focus of this article has been in the conformal
  spacetime set-up ---and hence restricted to trace-free matter models---
  an analogous calculation can be performed directly in the physical
  spacetime, lifting then the trace-free  matter restriction,
  obtaining a quadrupole-like formula for perturbations of the Weyl
  tensor.  Observe that, for the derivation of the wave equation
  satisfied by the rescaled Weyl tensor, the specific form of the
  rescaled Cotton tensor (in terms of the unphysical energy-momentum
  tensor) is never used. It follows that for the physical set-up, one 
  gets a
  formally identical equation.  The wave equation for the Weyl tensor
  has been derived for instance in \cite{Hol22} and \cite{Bie14}.
  Therefore, the analysis of section \ref{sec:quad_Weyl} would still
  hold replacing the rescaled Weyl tensor $d_{abcd}$ with the Weyl
  tensor $C_{abcd}$ and the rescaled Cotton tensor $T_{abc}$ by its
  physical counterpart $\tilde{T}_{abc}$ which written in terms of the
  physical energy-momentum reads:
  \begin{align}
    \tilde{T}_{abc}:= \tilde{\nabla}_{[a}\tilde{T}_{b]c} +
    \frac{1}{3}\tilde{g}_{c[b}\tilde{\nabla}_{a]}(\lambda- \tilde{T}).
  \end{align}}
\end{IndRemark}
\begin{IndRemark} \label{Rem:Gauge}
  \emph{ We note that the expression for $S_{\mu\nu\alpha\beta}$ is
  independent of the choices for the gauge source function. For this
  reason, one can choose the scri-fixing gauge independently of the
  assumption that $S_{\mu\nu\alpha\beta}$ has compact support in
  $\Omega$; the quadrupole-like formula we obtain for the rescaled Weyl
  tensor is therefore compatible with any gauge choice, in particular
  the scri-fixing gauge introduced earlier.}
\end{IndRemark}

\section{Summary and discussion}

In this article, we have presented the linearisation of the wave-like
hyperbolic reduction of the CEFEs in a GHG gauge for a general
background. As a first step toward studying the linear perturbations of
spacetimes using the conformal CEFE approach, the equations have been
particularised to the case where the background is the inversion
(unphysical) Minkowski spacetime. Conceptually, when studying
perturbations of the physical Minkowski spacetime, one can in principle
use any conformally flat metric so that
$\mathring{d}_{\mu\nu\alpha\beta}=0$. However having a non-vanishing
$\mathring{\Phi}_{\mu\nu}$ makes the equations quite cumbersome. Hence,
as a first analysis, we have opted to use as a background the inversion
Minkowski spacetime for which $\mathring{\Phi}_{\mu\nu}=\mathring{R}=0$.
For this background, we discuss the propagation of the gauge (the
generalised Lorenz gauge and the linear version of the conformal
constraints). The fact that the chosen background is flat permits the
use of techniques similar to those of the classical methods in physical
spacetime to study gravitational waves. In this manner, we have obtained
the conformal CEFE counterpart of the classical linearisation of the
(physical) Einstein field equations.

Nevertheless, one may wish to consider linearising around non-flat
conformal representations of the Minkowski spacetime. A particularly
interesting choice is the $i^0$-cylinder representation of Minkowski
spacetime, as this is particularly useful for studying the gravitational
field close to the critical sets $\mathcal{I}^\pm$---the region where
null-infinity $\mathscr{I}$ and spatial infinity $i^0$ meet.  Although
this analysis has been done via the spin-2 equation
$\nabla_{A'}{}^{A}\phi_{ABCD}=0$ in \cite{Val03a} (see also
\cite{MinMacVal22, DuaFenGasHil23} for a discussion of the solutions of
wave equations in this background) where it has been shown that the
(linearisation of rescaled Weyl spinor) spin-2 field develops
logarithmic terms, it would be of interest to see how these logarithmic
terms manifest in the metric, which is the variable employed in most
perturbation schemes in physics (such as the post-Minkowskian
expansions). Another interesting class of backgrounds are Petrov type D
spacetimes such as Kerr, so that a conformal Teukolsky-like equation may
be obtained; however, a different linearisation procedure (such as that
of Chandrasekhar\cite{Cha83a}), may be more appropriate in that case.
These avenues will be pursued in future work.  More generally, the
present work represents a first step into studying linear perturbations
of spacetimes of physical interest (such as black holes) within the CEFE
framework. 

A particularly useful result of this paper is the development of a
scri-fixing strategy using gauge source functions for CEFEs.  Our
strategy provides an alternative to more abstract gauges based on
conformal curves and conformal geodesics \cite{LueVal12}, one that is
particularly well-suited for numerical implementation. Though the
strategy is initially developed for the linearised equations (which are
expected to be good approximation in the weak field region of
asymptotically flat spacetimes of physical interest), the generalisation
to the nonlinear case, as discussed in the main text, is
straightforward. A minor limitation of this scri-fixing strategy in the
asymptotically flat setting is that the initial data must be prescribed
on slices that intersect future null infinity (as opposed to spatial
infinity), but our scri-fixing strategy fits well within the
hyperboloidal approach to Numerical Relativity.

Finally, we have constructed quadrupole-like formulae for the CEFEs
linearised around a flat background. While the quadrupole-like formulae
for $\delta g_{\mu \nu}$ and $\delta \Phi_{\mu \nu}$ require gauge
source functions with compact support and a trace-free energy-momentum
tensor, the quadrupole-like formula we obtain for the Weyl tensor is
independent of gauge choice and is formally identical to the case where
the energy-momentum tensor has nonvanishing trace. Since the Weyl tensor
yields an unambigious notion of gravitational radiation at future null
infinity, our result provides a familiar prescription for computing
gravitational waveforms at null infinity generated by weakly gravitating
sources in the linear approximation.

\subsection*{Acknowledgements}

We have profited from conversations with Juan Valiente Kroon and David
Hilditch and Alex Va\~n\'o Vi\~nuales.  We also thank J. Valiente
Kroon for sharing a Mathematica notebook where the main calculations
of \cite{CarHurVal} were done.  EG holds a FCT (Portugal) investigator
grant 2020.03845.CEECIND.  This paper is part of the Exploratory
Research Project 2022.01390.PTDC funded by FCT. JF acknowledges
support from FCT (Portugal) program UIDB/00099/2020. JF is grateful to
the Yukawa Institute for Theoretical Physics, Kyoto University, which
hosted a visit during which a part of this work was performed.

\appendix

\section{Explicit expressions for linearised CEFEs}
\label{Appendix:A}

Here, we provide expressions for the right-hand side of the linearised
wave-CEFEs, which are given below ($\bm\phi$ and $\bmf$ schematically
representing the respective field variables and gauge sources):
\begin{equation}\label{eq:appendix_wave_eqs}
  \begin{aligned}
    \mathring{\square} \delta g_{\mu \nu }
    &=
    H^{g}_{(\mu\nu)}(\delta \bm\phi,
    \dotnabla\delta\bm\phi \; ; \;  \mathring{\bm\phi}, \bmf, 
    \dotnabla\mathring{\bm\phi},
    \dotnabla \bmf )
    \\
    \mathring{\square} \delta  \Xi 
    & = H^{\Xi}_{}(\delta\bm\phi, \dotnabla \delta\bm\phi
    \; ; \; 
    \mathring{\bm\phi}, \bmf, \dotnabla\mathring{\bm\phi} )  \\
    \mathring{\square} \delta  s 
    & = 
    H^{s}(\delta\bm\phi, \dotnabla\delta\bm\phi
    \; ; \; 
    \mathring{\bm\phi}, \bmf, \dotnabla\mathring{\bm\phi},
    \dotnabla \bmf )   \\ 
    \mathring{\square} \delta \Phi_{\mu \nu }
    &= 
    H^{\Phi}_{(\mu\nu)}(\delta \bm\phi,
    \dotnabla \delta \bm\phi \; ; \; 
    \mathring{\bm\phi}, \bmf, \dotnabla\mathring{\bm\phi},
      \dotnabla \bmf,  \dotnabla\dotnabla \mathring{F}, \dotnabla\dotnabla \delta F )
    \\
    \mathring{\square} \delta  d_{\mu \nu \alpha \beta}
    & =
    H^{d}_{[\mu\nu][\alpha\beta]}(\delta \bm\phi, 
    \dotnabla \delta \bm\phi
    \; ; \; 
    \mathring{\bm\phi}, \bmf, \dotnabla\mathring{\bm\phi},
    \dotnabla\dotnabla\mathring{\bm\phi}, \dotnabla \bmf ) .
  \end{aligned}
 \end{equation}

To give more concise expressions, the quantities $\bmH$ on the
right-hand side of the equations above are provided below without
explicit (anti)symmetrisation; the appropriate (anti)symmetrisation of
indices is indicated in equation \eqref{eq:appendix_wave_eqs}.  This
calculation has been performed using the {\tt{xPert}} package of the
suite {\tt{xAct}}  in {\tt{Mathematica}} \cite{xAct_web} for metric
perturbation theory.

\begin{align}
  H^{g}_{\mu\nu} &:= -4 \delta \Phi _{\mu \nu } +8 \dotPhi
  _{(\mu }{}^{\alpha }\delta g_{\nu )\alpha } -2 \dotXi \delta
  g^{\alpha \beta }\mathring{d}_{(\mu |\alpha |\nu )\beta } 
  + \tfrac{1}{6} \dotF \delta
  g_{\mu \nu } + \delta g^{\alpha }{}_{\alpha } (-2 \dotPhi _{\mu \nu }
  - \tfrac{1}{6} \dotF \dotg_{\mu \nu })
  + 2 \dotnabla _{(\mu }\dotF_{\nu )}
  \nonumber \\ &
  + \dotg_{\mu \nu }
  (\tfrac{1}{2} \delta F + 2 \delta \Phi ^{\alpha }{}_{\alpha } -2
  \dotPhi^{\alpha \beta } \delta g_{\alpha \beta } - \dotnabla
  _{\alpha }\dotF^{\alpha })
\end{align}

\begin{flalign}
  H^{\Xi}_{} & := - \tfrac{1}{6} \dotF \delta \Xi - \tfrac{1}{6}
  \dotXi \delta F + 4 \delta s + (\tfrac{1}{2} \dotXi ^3 \dotT^{\mu
    \nu } - \dotXi \dotPhi ^{\mu \nu }) \delta g_{\mu \nu } +
  (\tfrac{1}{24} \dotXi \dotF - \mathring{s}) \delta g^{\mu }{}_{\mu } +
  \dotF^{\mu } \dotnabla_{\mu }\dotXi
\end{flalign}

\begin{align}
  H^{s} & := \big(\tfrac{\dotF}{12}\big)^2 (\delta \Xi
  - \tfrac{1}{4}  \dotXi \delta g^{\mu }{}_{\mu })
  + \dotPhi ^{\mu \nu } (2 \dotXi  \delta \Phi _{\mu \nu }
  - \dotXi ^3 \delta T_{\mu \nu } 
  + \dotPhi_{\mu \nu } 
  (\delta \Xi + \tfrac{1}{2} \dotXi \delta g^{\alpha  }{}_{\alpha })
  + \tfrac{1}{12} \dotXi \dotF \delta g_{\mu \nu }
  -  \mathring{s} \delta g_{\mu \nu } \nonumber \\ & 
  - \dotXi \dotPhi_{\mu }{}^{\alpha } \delta g_{\nu \alpha })
  + \dotT^{\mu \nu } (- \dotXi ^3 \delta \Phi _{\mu    \nu } 
  + \tfrac{1}{2} \dotXi ^5 \delta T_{\mu \nu }
  -  \tfrac{1}{2} \dotXi ^2 \dotPhi _{\mu \nu } 
  (6 \delta \Xi + \dotXi  \delta g^{\alpha }{}_{\alpha })
  + \tfrac{1}{8} \dotXi ^4 \dotT_{\mu    \nu } 
  (10 \delta \Xi + \dotXi \delta g^{\alpha }{}_{\alpha }) \nonumber \\ &
  + \tfrac{1}{2} \dotXi ^2  \delta g_{\mu \nu }
  ( \mathring{s}  -  \tfrac{1}{12} \dotXi  \dotF )
  + ( \dotXi ^3 \dotPhi _{\mu }{}^{\alpha }
    - \tfrac{1}{4} \dotXi ^5  \dotT_{\mu }{}^{\alpha }
    )
  \delta g_{\nu \alpha }) +
  \tfrac{\dotF}{24} 
  (
    \mathring{s} \delta g^{\mu  }{}_{\mu } + \tfrac{1}{3} 
    (\dotXi \delta F  - 12 \delta s  
    - 3  \dotF^{\mu } \dotnabla _{\mu }\dotXi )
  )
  \nonumber \\ &
  + \tfrac{1}{6} \dotnabla ^{\mu }\dotXi (3 \delta  g^{\nu \alpha}
  (\dotXi ^2 \dotnabla _{\alpha }\dotT_{\mu \nu }
  -2  \dotnabla _{\alpha }\dotPhi _{\mu \nu }) 
  - \dotnabla _{\mu }\delta  F)  
  - \tfrac{1}{6} ( \mathring{s} \delta F  
  + \dotnabla_{\mu }\delta \Xi \dotnabla ^{\mu }\dotF)
  - \dotF^{\mu }  \dotPhi _{\mu \nu } \dotnabla ^{\nu }\dotXi 
  \nonumber \\ &
  + (\dotXi \delta T_{\mu    \nu }
  - \dotXi \dotT_{\mu }{}^{\alpha } \delta g_{\nu \alpha })
  \dotnabla ^{\mu }\dotXi \dotnabla ^{\nu }\dotXi 
  + \tfrac{1}{8}  \delta g_{\mu \nu }
  \dotnabla ^{\mu }\dotXi \dotnabla ^{\nu }\dotF +  \dotT_{\mu \nu }
  (\tfrac{1}{2} \dotXi ^2 \dotF^{\mu } \dotnabla^{\nu }\dotXi  
  + 2  \dotXi \dotnabla ^{\mu }\dotXi \dotnabla ^{\nu }\delta \Xi
  \nonumber\\ &
  + (\delta \Xi + \tfrac{1}{2} \dotXi \delta g^{\alpha  }{}_{\alpha })
  \dotnabla ^{\mu }\dotXi \dotnabla ^{\nu }\dotXi)
\end{align}

\begin{align}
  H^{\Phi}_{\mu\nu} & := \tfrac{1}{3} \dotF \delta \Phi _{\mu  \nu  }
  + \tfrac{1}{2} \dotXi ^3 \dotT^{\alpha  \delta  } 
    \delta d_{\mu  \alpha  \nu  \delta  }
  - 2 \dotPhi _{\mu  }{}^{\delta  } 
    \dotPhi _{\nu  \delta  } \delta g^{\alpha  }{}_{\alpha  }
  - \tfrac{1}{12} \dotPhi _{\mu  \nu  } 
    (2 \delta F  + \dotF \delta g^{\alpha  }{}_{\alpha  })
  - 2 \dotXi  \dotPhi _{\mu  }{}^{\alpha  } \delta g^{\delta  \beta  } 
    \dotd_{\nu  \delta  \alpha  \beta  }
  \nonumber \\ &
  + \dotd_{\mu  \alpha  \nu  \delta } 
  (-2 \dotPhi ^{\alpha  \delta  } \delta \Xi
  + \tfrac{1}{2} \dotXi  (3 \dotXi  \dotT^{\alpha  \delta  } \delta \Xi
  - 4 \delta \Phi ^{\alpha  \delta  } 
  + \dotXi ^2 \delta T^{\alpha  \delta  }))
  + \tfrac{1}{24} (\delta g^{\alpha  }{}_{\alpha  } \dotg_{\mu  \nu  }
  - \tfrac{1}{2} \delta g_{\mu  \nu  }) \square _{\dotg}{} \dotF
  \nonumber \\ &
  + (4 \dotXi  \dotPhi ^{\alpha  \delta  } 
    \delta g_{\alpha  }{}^{\beta  }
    - \dotXi  \dotPhi^{\delta  \beta  } \delta g^{\alpha  }{}_{\alpha  }
    + \tfrac{1}{4} \dotXi ^3 (
        -4 \dotT^{\alpha  \delta  } \delta g_{\alpha  }{}^{\beta  }
        + \dotT^{\delta  \beta  } \delta g^{\alpha  }{}_{\alpha  })
    ) \dotd_{\mu  \delta  \nu  \beta  }
  - 2 \delta T_{\mu  }{}^{\alpha  }{}_{\nu  } 
    \dotnabla _{\alpha  }\dotXi
    \nonumber \\ &
  + \dotPhi ^{\alpha  \delta  } (
    - 2 \dotXi  \delta d_{\mu  \alpha  \nu  \delta  }
    - 2 \dotPhi _{\mu  \nu  } \delta g_{\alpha  \delta  }
    - \dotPhi _{\alpha  \delta  } \delta g_{\mu  \nu  }
    - 2 \delta \Phi _{\alpha  \delta  } \dotg_{\mu  \nu  }
    + 2 \dotPhi _{\alpha  }{}^{\beta  } \delta g_{\delta  \beta  } 
      \dotg_{\mu  \nu  })
  - 2 \dotT_{\mu  }{}^{\alpha  }{}_{\nu  } 
    \dotnabla _{\alpha  }\delta \Xi
  \nonumber \\ &
   + \dotF^{\alpha  } (\dotXi  \dotT_{\mu  \alpha  \nu  } 
   + \dotnabla _{\alpha  }\dotPhi _{\mu  \nu  })
   + \dotXi  \delta g^{\alpha  \delta  } \dotnabla _{\alpha  }
     \dotT_{\mu  \delta  \nu  }
   + \dotnabla _{\alpha  }\delta g_{\nu  \delta  } (
      - \dotXi  \dotT^{\alpha  \delta  }{}_{\mu  }
      + \tfrac{1}{2} \dotXi  \dotT_{\mu  }{}^{\alpha  \delta  }
      + 2 \dotnabla ^{\alpha  }\dotPhi _{\mu  }{}^{\delta  })
  \nonumber \\ &
    - \dotXi  \dotnabla ^{\alpha  }\delta T_{\mu  \alpha  \nu  }
  - \delta \Xi  \dotnabla ^{\alpha  }\dotT_{\mu  \alpha  \nu  }
  + 2 \delta g^{\alpha  \delta  } \dotT_{\mu  \alpha  \nu  } 
    \dotnabla _{\delta  }\dotXi
  + (- \tfrac{1}{2} \dotXi  \dotT_{\mu  }{}^{\alpha  \delta  }
    -2 \dotnabla ^{\alpha  }\dotPhi _{\mu  }{}^{\delta  }) 
    \dotnabla _{\delta  }\delta g_{\nu  \alpha  }
  \nonumber \\ &
  + \tfrac{1}{12} \dotnabla _{\alpha  }\delta g_{\mu  \nu  } 
    \dotnabla ^{\alpha  }\dotF
  + \tfrac{1}{24}\dotg_{\mu\nu}(
    - \dotnabla _{\alpha}\delta g^{\delta  }{}_{\delta}
      \dotnabla^{\alpha }\dotF
    + \dotF^{\alpha  } \dotnabla _{\alpha  }\dotF
    + \delta g^{\alpha  \delta  } \dotnabla _{\delta  }
      \dotnabla _{\alpha  }\dotF)
  + \delta g^{\alpha  \delta  } \dotnabla _{\delta  }
  \dotnabla _{\alpha  }\dotPhi _{\mu  \nu  }
  \nonumber \\ &
  + \dotnabla _{\alpha}\delta g^{\delta}{}_{\delta} (
    - \tfrac{3}{4} \dotXi \dotT_{\mu  }{}^{\alpha }{}_{\nu }
    - \dotnabla ^{\alpha  }\dotPhi _{\mu  \nu  } 
    + \dotnabla _{\mu  }\dotPhi _{\nu  }{}^{\alpha  })
  - \tfrac{1}{6} \dotnabla ^{\alpha  }\dotF \dotnabla _{\mu  }
    \delta g_{\nu  \alpha  }
  - ( \tfrac{1}{4} \dotXi  \dotT_{\mu  }{}^{\alpha  }{}_{\alpha  }
  + \dotnabla ^{\alpha  }\dotPhi _{\mu  \alpha  }) \dotnabla _{\nu  }
    \delta g^{\delta  }{}_{\delta }
  \nonumber \\ &
  + \dotPhi _{\mu  }{}^{\alpha  } (4 \delta \Phi _{\nu  \alpha  }
  + 2 \dotPhi _{\nu  }{}^{\delta  } \delta g_{\alpha  \delta  }
  + \tfrac{1}{6} \dotF \delta g_{\nu  \alpha  } 
  + 2 \dotPhi _{\alpha  }{}^{\delta  } \delta g_{\nu  \delta  }
  + 2 \dotnabla _{\nu  }\dotF_{\alpha  }) 
  + (\tfrac{1}{2} \dotXi  \dotT_{\mu  }{}^{\alpha  \delta  }
  + 2 \dotnabla ^{\alpha  }\dotPhi _{\mu  }{}^{\delta  }) 
    \dotnabla _{\nu  }\delta g_{\alpha  \delta  }
  \nonumber\\ &
  + \tfrac{1}{12} \dotnabla _{\mu  }\dotF \dotnabla _{\nu  }
    \delta g^{\alpha  }{}_{\alpha  }
  - \tfrac{1}{12} \delta g^{\alpha  }{}_{\alpha  } 
    \dotnabla _{\nu  }\dotnabla _{\mu  }\dotF
  + \tfrac{1}{6} \dotnabla _{\{\nu  }\dotnabla _{\mu\}  }\delta F
\end{align}

\begin{align}
  H^{d}_{\mu\nu\alpha\beta} & :=
  \tfrac{1}{2} \dotF \delta d_{\mu  \nu  \alpha  \beta  }
  - \dotXi^2 \dotT_{\alpha  }{}^{\delta  } 
    \delta d_{\mu  \nu  \beta  \delta  }
  - \dotXi ^2 \dotT_{\mu  }{}^{\delta  } 
    \delta d_{\nu  \delta  \alpha  \beta  }
  - ( \tfrac{1}{2} \delta F  
    + 4 \dotPhi ^{\delta  \sigma  } \delta g_{\delta  \sigma  }
    + \tfrac{1}{12} \dotF \delta g^{\delta  }{}_{\delta  }) 
    \dotd_{\mu  \nu  \alpha  \beta  }
  \nonumber \\ &
  + (-2 \dotXi  \dotT_{\alpha  }{}^{\delta  } \delta \Xi  
    + 4 \delta \Phi _{\alpha  }{}^{\delta  }
    - \dotXi ^2 \delta T_{\alpha  }{}^{\delta  }  
    - \tfrac{1}{6} \dotF \delta g_{\alpha  }{}^{\delta  }
    + 2 \dotPhi _{\alpha  }{}^{\delta  } 
    \delta g^{\sigma  }{}_{\sigma  }) \dotd_{\mu  \nu  \beta  \delta  }
  - 2 \dotXi  \delta d_{\alpha  }{}^{\delta  }{}_{\beta  }{}^{\sigma  } 
    \dotd_{\mu  \nu  \delta  \sigma  }
  \nonumber \\ &
  + (-2 \dotPhi ^{\delta  \sigma  } \delta g_{\alpha  \delta  }
    + \dotXi ^2 \dotT_{\alpha  }{}^{\delta  } 
    \delta g_{\delta  }{}^{\sigma  }
    - 6 \dotPhi _{\alpha  }{}^{\delta  } 
      \delta g_{\delta  }{}^{\sigma  })
    \dotd_{\mu  \nu  \beta  \sigma  }
  - 2\dotd_{\alpha  \delta  \beta  \sigma  } 
    ( \dotXi  \delta d_{\mu  \nu  }{}^{\delta  \sigma  }
    +  \delta \Xi  \dotd_{\mu  \nu  }{}^{\delta  \sigma  })
  \nonumber \\ &
  - (2 \dotXi  \dotT_{\mu  }{}^{\delta  } \delta \Xi
    - 4 \delta \Phi _{\mu  }{}^{\delta  }  
    + \dotXi ^2 \delta T_{\mu  }{}^{\delta  }
    + \tfrac{1}{6} \dotF \delta g_{\mu  }{}^{\delta  }
    - 2 \dotPhi _{\mu }{}^{\delta  } \delta g^{\sigma  }{}_{\sigma  }) 
    \dotd_{\nu  \delta  \alpha  \beta  }
  + 4 \delta g_{\mu  \alpha  } 
    \dotnabla _{\delta  }\dotT_{\beta  }{}^{\delta  }{}_{\nu  }
  + \dotT_{\mu  \alpha  \beta  } 
    \dotnabla _{\nu  }\delta g^{\delta  }{}_{\delta  }
  \nonumber \\ &
  - 2(4 \dotXi  \delta d_{\mu  }{}^{\delta }{}_{\alpha  }{}^{\sigma  }
    + \dotXi ^2 \dotT^{\delta  \sigma  } \delta g_{\mu  \alpha  }
    + 2 \delta \Xi  
      \dotd_{\mu  }{}^{\delta  }{}_{\alpha  }{}^{\sigma  }) 
    \dotd_{\nu  \delta  \beta  \sigma  }
  + (\dotXi ^2 \dotT_{\mu }{}^{\delta  } 
    \delta g_{\delta  }{}^{\sigma  }
    - 6 \dotPhi _{\mu  }{}^{\delta  } \delta g_{\delta  }{}^{\sigma  }
    -2 \dotPhi ^{\delta  \sigma  } \delta g_{\mu  \delta  }) 
    \dotd_{\nu  \sigma  \alpha  \beta  }
  \nonumber \\ &
  + 2\dotXi \delta g^{\delta \sigma} (
    \dotd_{\beta \delta \sigma \gamma}\dotd_{\mu\nu\alpha}{}^{\gamma}
    + \dotd_{\mu  }{}^{\gamma}{}_{\alpha  \beta  } 
    \dotd_{\nu  \delta  \sigma  \gamma})
  + 4 \dotXi  \delta g^{\delta \sigma } (
    \dotd_{\alpha \sigma \beta  \gamma} 
    \dotd_{\mu  \nu  \delta }{}^{\gamma}
    + \dotd_{\mu  }{}^{\gamma}{}_{\alpha  \delta  } 
    \dotd_{\nu  \gamma \beta  \sigma  } 
    + \dotd_{\mu  \delta  \alpha  }{}^{\gamma} 
    \dotd_{\nu  \sigma  \beta  \gamma}) 
  \nonumber \\ &
  - \dotXi  \delta g^{\delta  }{}_{\delta  }
  ( \dotd_{\alpha  \sigma  \beta  \gamma} 
    \dotd_{\mu  \nu  }{}^{\sigma  \gamma}
  + 2 \dotd_{\mu  }{}^{\sigma  }{}_{\alpha  }{}^{\gamma} 
    \dotd_{\nu  \sigma  \beta  g})
  - 2 \dotXi  (2 \dotT^{\delta  \sigma  } \delta \Xi
    + \dotXi  \delta T^{\delta  \sigma  }) 
    \dotd_{\mu  \delta  \alpha  \sigma  } 
    \dotg_{\nu  \beta  }
  + \mathring{F} ^{\delta  } \dotnabla _{\delta  }
    \dotd_{\mu  \nu  \alpha  \beta  }
  \nonumber \\ &
  + \dotT^{\delta \sigma }\delta g_{\delta }{}^{\gamma}
    (2\dotXi^2\dotd_{\mu \gamma \alpha \sigma }\dotg_{\nu\beta}  
    + 2 \dotXi ^2 \dotd_{\mu  \sigma  \alpha  \gamma} 
      \dotg_{\nu  \beta  })
  + 2 \dotnabla _{\alpha  }\delta T_{\mu  \nu  \beta  }
  + 2 \dotd_{\mu  \nu  \alpha  \delta  } \dotnabla _{\beta  }
    \dotF^{\delta  }
  - \dotT_{\mu  \alpha  \nu  } \dotnabla _{\beta  }
    \delta g^{\delta  }{}_{\delta  }
  \nonumber \\ &
  + \dotg_{\nu  \beta  } 
    (-2 (\dotXi ^2 \dotT^{\delta  \sigma  } 
        \delta d_{\mu  \delta  \alpha  \sigma  }
        + 2 \dotF^{\delta  } \dotT_{\alpha  \delta  \mu  })
    + (4 \dotT_{\alpha  }{}^{\delta  }{}_{\mu  } 
      + \dotT_{\mu  \alpha  }{}^{\delta  }
      + \dotT_{\mu  }{}^{\delta  }{}_{\alpha  }) 
      \dotnabla _{\delta  }\delta g^{\sigma  }{}_{\sigma  })
  - \dotg_{\mu\beta}\dotT_{\alpha  }{}^{\delta  }{}_{\delta  } 
    \dotnabla _{\nu  }\delta g^{\sigma  }{}_{\sigma  }
  \nonumber \\ &
  + 2 \dotnabla _{\mu  }\delta T_{\alpha  \beta  \nu  }
  - \delta g^{\delta  }{}_{\delta  } 
    ( \dotnabla _{\alpha  }\dotT_{\mu  \nu  \beta  }
    + \dotnabla _{\mu  }\dotT_{\alpha  \beta  \nu  })
  + 2 \dotd_{\mu  \delta  \alpha  \beta  } \dotnabla _{\nu  }
    \dotF^{\delta  }
  + 4 \dotT_{(\alpha  }{}^{\delta  }{}_{\mu ) }
    ( \dotnabla _{\beta  }\delta g_{ \delta\nu  } 
    + \dotnabla _{\nu  }\delta g_{\beta  \delta  }
    -  \dotnabla _{\delta  }\delta g_{\beta  \nu  })
    \nonumber \\ &
  + 4 \dotT^{\delta  \sigma  }{}_{\mu  } \dotg_{\nu  \alpha  } 
    \dotnabla _{\sigma  }\delta g_{\beta  \delta  }
  + 2 \dotg_{\mu  \beta  }\dotT_{\alpha  }{}^{\delta  \sigma  }
    ( \dotnabla _{\delta  }\delta g_{\nu  \sigma  }
    + \dotnabla _{\nu  }\delta g_{\delta  \sigma  }
    - \dotnabla _{\sigma  }\delta g_{\nu  \delta  })
    + 4 \dotnabla _{[\delta  }
      \dotd_{|\mu  \nu  \beta|  \sigma]}
      \dotnabla ^{\sigma  }\delta g_{\alpha  }{}^{\delta }
    \nonumber \\ &
  + \dotg_{\mu  \alpha  } 
    (4 \dotnabla _{\delta  }\delta T_{\beta  }{}^{\delta  }{}_{\nu  }
    - 4 \delta g^{\delta  \sigma  } \dotnabla _{\sigma  }
      \dotT_{\beta  \delta  \nu  }
    - 2 \delta g^{\delta  }{}_{\delta  } \dotnabla _{\sigma  }
      \dotT_{\beta  }{}^{\sigma  }{}_{\nu  })
  - 2 \dotnabla _{\alpha  }\delta g^{\delta  \sigma  } 
    \dotnabla _{\sigma  }\dotd_{\mu  \nu  \beta  \delta  }
  - 2 \dotnabla _{\mu  }\delta g^{\delta  \sigma  } 
    \dotnabla _{\sigma  }\dotd_{\nu  \delta  \alpha  \beta  }
  \nonumber \\ &
  + 8 \dotnabla _{[\sigma  }\dotd_{ \delta] \nu  \alpha  \beta}
    \dotnabla ^{\sigma  }\delta g_{\mu  }{}^{\delta  }
  - 2 ( \dotnabla _{[\alpha  }\dotd_{|\mu \nu \beta| \sigma] } 
      + \dotnabla _{[\mu  }\dotd_{|\nu| \sigma]\alpha\beta } ) 
      \dotnabla ^{\sigma }\delta g^{\delta }{}_{\delta }
  + \delta g^{\delta  \sigma  } \dotnabla _{\sigma  }
    \dotnabla _{\delta  }\dotd_{\mu  \nu  \alpha  \beta  }
  \end{align}

\section{Reduced wave equations}
\label{Appendix:B}

In this appendix, we extend the analysis leading up to
\eqref{reduced_wave_metric} to the other wave equations listed in
\eqref{geo_wave} for the sake of completeness. In particular, we wish to
recast \eqref{geo_wave} explicitly as a system of second order
hyperbolic equations.  This non-linear system of wave equations was
originally derived in \cite{CarHurVal} using slightly different set of
variables and a different reduced wave operator.

\medskip

First observe that in general, the geometric wave operator acting on a
tensor will give rise to derivatives of the Christoffel symbols. Since
the metric is a variable of the system, this is problematic for
hyperbolicity. In particular, for a covector $w_a$ one has
\begin{align}\label{BoxTransVect}
  \square w_\mu=  \reducedbox w_{\mu  }
  - w^{\nu  } \nabla _{\alpha  }\Gamma_{\nu  \mu  }{}^{\alpha  }
  -2 \Gamma_{\nu  \mu  \alpha  } \nabla ^{\alpha  }w^{\nu  }
  - \Gamma _{\nu  } \nabla ^{\nu  }w_{\mu  }
  - \Gamma^{\alpha  }{}_{\mu  }{}^{\beta  } 
    \Gamma_{\nu  \alpha  \beta  } w^{\nu  }
  - \Gamma_{\nu  \mu  }{}^{\alpha  } \Gamma _{\alpha  } w^{\nu  }.
\end{align}
In this appendix, indices are raised and lowered with the metric $g_{\mu
\nu}$. Observe that the second term in the right-hand side of equation
$\eqref{BoxTransVect}$ represents second derivatives of the metric.
However, one can replace these derivatives of the Christoffel symbols by
noticing the identity
\begin{align}\label{Key_id}
  g^{\mu \alpha } \nabla _{\alpha }\Gamma ^{\beta }{}_{\nu \mu } = -
  R_{\nu }{}^{\beta }+\Gamma^{\beta \mu \alpha } \Gamma _{\mu \nu
    \alpha } - \Gamma^{\beta }{}_{\nu \mu } \Gamma ^{\mu } + \nabla
  _{\nu }\Gamma^{\beta }.
\end{align}
Exploiting the identity \eqref{Key_id} a direct calculation gives
\begin{align}
  \square w_{\mu  } =  \reducedbox w_{\mu  } +
  \frac{1}{4} R w_{\mu } + 2 \Phi _{\mu  \nu  } w^{\nu  }
  - \mathcal{L}_\Gamma  w_{\mu  }
  -2 \Gamma_{\nu  \mu  \alpha  } \nabla ^{\alpha  }w^{\nu  }
  -2 \Gamma^{\alpha  }{}_{\mu  }{}^{\beta  } 
    \Gamma_{\nu  \alpha  \beta  } w^{\nu  }  
\end{align}
where $\mathcal{L}_\Gamma$ is the Lie-derivative along $\Gamma^a$. Using
the definition of the conformal and the GHG-constraints one can rewrite
the last expression as
\begin{align}
  \square  w_{\mu  } =  \reducedbox w_{\mu  } +
  \frac{1}{4} C w_{\mu } - \mathcal{L}_C  w_{\mu  } + 
  \frac{1}{4} F w_{\mu }
  + \mathcal{L}_{\mathcal{H}}  w_{\mu  } + \mathcal{Q}(w)_\mu
\end{align}
where $\mathcal{L}_C$ and $\mathcal{L}_{\mathcal{H}}$ are the Lie
derivative along $C^a$ and $\mathcal{H}^a$ respectively and
\begin{align}
  \mathcal{Q}(w)_\mu := 2 \Phi _{\mu  \nu  } w^{\nu  }
  -2 \Gamma_{\nu  \mu  \alpha  } \nabla ^{\alpha  }w^{\nu  }
  -2 \Gamma^{\alpha  }{}_{\mu  }{}^{\beta  } 
  \Gamma_{\nu  \alpha  \beta  } w^{\nu  }  
\end{align}
One can generalise the latter calculation for tensor of type $(0,n)$ as
follows
\begin{align}
  \square  w_{\mu_{1} ...\mu_{n} } = \reducedbox w_{\mu_{1}
    ...\mu_{n}} + \frac{n}{4} C w_{\mu } - \mathcal{L}_C w_{\mu_{1}
    ...\mu_{n}} + \frac{n}{4} F w_{\mu_{1} ...\mu_{n}} 
    + \mathcal{L}_{\mathcal{H}}
  w_{\mu_{1} ...\mu_{n}} + \mathcal{Q}(w)_{\mu_{1} ...\mu_{n}}
\end{align}
where
\begin{align}
  \mathcal{Q}(w)_{\mu_{1} ...\mu_{n} } := & 2 \Phi _{\mu_{1} \nu }
  w^{\nu}{}_{\mu_2...\mu_{n}} -2 \Gamma_{\nu \mu_{1} \alpha } \nabla
  ^{\alpha }w^{\nu}{}_{\mu_2...\mu_n} -2 \Gamma^{\alpha }{}_{\mu_{1}
  }{}^{\beta } \Gamma_{\nu \alpha \beta } w^{\nu}{}_{\mu_2...\mu_n} +
  ...+ \\ & + 2 \Phi _{\mu_{n} \nu } w_{\mu_{1}...\nu_{n-1}}^{\nu} -2
  \Gamma_{\nu \mu_{n} \alpha } \nabla ^{\alpha
  }w_{\mu_1...\mu_{n-1}}{}^\nu -2 \Gamma^{\alpha }{}_{\mu_{n}
  }{}^{\beta } \Gamma_{\nu \alpha \beta } w_{\mu_1...\mu_{n-1}}{}^\nu.
\end{align}
Then, by imposing the conformal and GHG-constraints are satisfied so
that $C=0$ and $C^a=0$ one effectively has effectively removed the
problematic terms for hyperbolicity:
\begin{align}
  \square w_{\mu_{1} ...\mu_{n} } = \reducedbox w_{\mu_{1} ...\mu_{n}}
  + \frac{n}{4} F w_{\mu_{1} ...\mu_{n}} + \mathcal{L}_{\mathcal{H}}
  w_{\mu_{1} ...\mu_{n}} + \mathcal{Q}(w)_{\mu_{1} ...\mu_{n}}
\end{align}
Applying the latter rule on each of the fields appearing on the
geometric wave equations one obtains a set of hyperbolic non-linear
wave equations which read as follows.

\bigskip

\noindent For the components of the metric:

\begin{subequations}\label{red_wave_CEFEs}

\begin{align}\label{reduced_wave_metric_0}
\reducedbox\; g_{\mu \nu } = - 4\Phi_{\mu\nu} 
- \tfrac{1}{2}F g_{\mu\nu} -
2 \nabla _{(\mu }\mathcal{H}_{\nu) } - g^{\sigma\delta }g^{\alpha \beta}
(\Gamma_{\nu \delta \beta }\Gamma_{\sigma \mu \alpha } +
2\Gamma_{\delta \nu \beta }\Gamma_{(\mu\nu) \alpha } ).
\end{align}

\noindent For the components of the trace-free Ricci tensor:

\begin{align}\label{reduced_wave_Phi}
  & \reducedbox \Phi_{\mu\nu} = - \tfrac{1}{6} \Phi _{\mu\nu} F
  -\mathcal{L}_{\mathcal{H}}\Phi_{\mu\nu} - \mathcal{Q}(\Phi)_{\mu\nu} 
    + \tfrac{1}{6} \nabla _{\{\mu}\nabla _{\nu\}}F + 
  4 \Phi _{\mu}{}^{\alpha} \Phi _{\nu\alpha}
  - \Phi _{\alpha\beta} \Phi ^{\alpha\beta} g_{\mu\nu}
  -2 \Xi  \Phi ^{\alpha\beta} d_{\mu\alpha\nu\beta}
   \nonumber
  \\  & \qquad \qquad \qquad \qquad   \qquad \qquad \qquad \qquad \qquad
  +   \tfrac{1}{2} \Xi ^3 T^{\beta \alpha} d_{\mu\beta\nu\alpha}
  - \Xi  \nabla _{\alpha}T_{\nu}{}^{\alpha}{}_{\mu}
  - 2T_{(\mu|\alpha|\nu)} \nabla ^{\alpha}\Xi.
\end{align}

\noindent For the components of the rescaled Weyl tensor:
\begin{align}\label{reduced_wave_rescaledWeyl}
  \reducedbox d_{\mu\nu\alpha\beta}  
  =& -\tfrac{1}{2}F d_{\mu\nu\alpha\beta}
  -\mathcal{L}_{\mathcal{H}}d_{\mu\nu\alpha\beta} 
  -\mathcal{Q}(d)_{\mu\nu\alpha\beta} + 2
  \Xi d_{\mu}{}^{\sigma}{}_{\beta}{}^{\lambda}
  d_{\nu\sigma\alpha\lambda} -2 \Xi
  d_{\mu}{}^{\sigma}{}_{\alpha}{}^{\lambda} d_{\nu\sigma\beta\lambda}
  -2 \Xi d_{\mu\nu}{}^{\sigma\lambda} d_{\alpha\sigma\beta\lambda}
  \nonumber \\ & - \Xi^2T_{[\alpha}{}^{\sigma}d_{\beta]\sigma\mu\nu} -
  \Xi^2T_{[\mu}{}^{\sigma}d_{\nu]\sigma\alpha\beta} +
  \Xi^2T^{\sigma\lambda}( d_{\mu\sigma \lambda[\alpha}g_{\beta]\nu} -
  d_{\nu\sigma\lambda[\alpha}g_{\beta]\mu} ) + 2
  \nabla_{[\mu}T_{|\alpha\beta|\nu]}  \nonumber \\ & 
  \qquad \qquad \qquad \qquad \qquad \quad \;
  +2  \nabla_{[\alpha}T_{|\mu\nu|\beta]} + 2
  g_{\beta[\nu}\nabla_{|\sigma}T_{\alpha|}{}^{\sigma}{}_{\mu]} + 2
  g_{\alpha[\mu}\nabla_{|\sigma}T_{\beta|}{}^{\sigma}{}_{\nu]}.
 \end{align}

\noindent For the Friedrich scalar:

\begin{align}\label{reduced_wave_s}
  \reducedbox s = &   - \tfrac{1}{6} F s 
  - \mathcal{L}_{\mathcal{H}} s   
  - \tfrac{1}{6} g_{\mu  \nu  } \nabla ^{\mu  }\Xi  \nabla ^{\nu  }F
  + \big(\tfrac{F}{12}\big)^2 \Xi 
  + \Xi  \Phi _{\mu  \nu } \Phi ^{\mu  \nu  }
  + \tfrac{1}{4} \Xi^5 T_{\mu\nu} T^{\mu\nu}
  - \Xi ^3 T^{\mu\nu} \Phi _{\mu\nu}
  \nonumber \\ &
  \qquad\qquad\qquad\qquad\qquad\qquad\qquad\qquad\qquad\qquad\qquad
  \qquad\qquad\quad
  + \Xi  T_{\mu\nu} \nabla ^{\mu}\Xi  \nabla ^{\nu}\Xi.
\end{align}

\noindent For the conformal factor:

\begin{align}\label{reduced_wave_conf_factor}
  \reducedbox \; \Xi = - \tfrac{1}{6} \Xi F
  - \mathcal{L}_{\mathcal{H}}\Xi   + 4 s .
\end{align}
\end{subequations}
\begin{IndRemark}
  \emph{ In vacuum, the hyperbolicity of the wave equations
  \eqref{red_wave_CEFEs} is clear. However, for the case with matter
  $T_{ab}=T_{ab}(\bm\tau, \nabla \bm\tau)$, one could at first instance
  question their hyperbolicity due to the presence of first derivatives
  of the rescaled Cotton tensor $T_{abc}$ which encode second
  derivatives of the energy-momentum tensor $T_{ab}$ ---hence the
  equations contain two (or more) derivatives of the matter fields
  $\bm\tau$. Nonetheless, one can still ensure hyperbolicity
  constructing wave equations for $\bm\tau$ and $\nabla \bm\tau$ so that
  $T_{abc}$ is expressed in terms of evolved fields only. For instance,
  in the case of the conformally invariant scalar field
  $T_{ab}=T_{ab}(\phi, \nabla\phi)$  one needs to construct wave
  equations for $\phi$ and a reduction variable $\varphi=\nabla\phi$
  ---see \cite{CarHurVal}.  Since this is a case-dependent analysis and
  this has been done in \cite{CarHurVal} for a number of matter models,
  further discussion is omitted. \emph}
\end{IndRemark}

In \cite{CarHurVal} the reduced wave operator is defined so that the
lower order contributions (along with terms involving the gauge source
functions) are absorbed into the definition of the operator
$\blacksquare$ in such a way that the reduced equations look formally
identical to the geometric ones. Here the standard reduced wave operator
$\reducedbox:=g^{\alpha\beta}\partial_\alpha\partial_\beta$ used in most
discussions of generalised harmonic gauge in Numerical Relativity was
employed instead. The initial data for the reduced wave equations
\eqref{red_wave_CEFEs} have to satisfy
\begin{equation}\label{vanishing_CFEs_tensorial_zq_ID}
  Z_{\mu\nu}|_{\Sigma}=0, \quad Z_{\mu}|_{\Sigma}=0, \quad
  \delta_{\mu\nu\sigma}|_{\Sigma}=0, \quad
  \lambda_{\mu\nu\sigma}|_{\Sigma}=0, \quad Z|_{\Sigma}=0.
\end{equation}
where $\Sigma \subset \mathcal{M}$ is a spacelike hypersurface on which
the initial data is prescribed. It should also be noticed that the
GHG-constraint $C^{\mu}$ and the conformal constraint $C$ are not
independent, in fact a calculation shows $C=\nabla_\mu C^{\mu}$.
Propagation of the gauge and a full discussion of the propagation of the
constraints [i.e. propagation of the zero-quantities
\eqref{vanishing_CFEs_tensorial_zq_ID}] has carefully been done in
\cite{CarHurVal} and will not be reproduced here.  The aim of revisiting
the derivation of the reduced wave equations \eqref{red_wave_CEFEs} is
simply to provide context and the \emph{non-linear analogue} of the
linearised equations derived in Section \ref{sec:lin_wave_CEFE_gen}.

\bibliographystyle{unsrt}

\begin{thebibliography}{10}

\bibitem{HilNur16}
C~Denson Hill and Pawel Nurowski.
\newblock How the green light was given for gravitational wave search, 2016.

\bibitem{BieGarYun17}
Lydia Bieri, David Garfinkle, and Nicolas Yunes.
\newblock Gravitational waves and their mathematics.
\newblock 2017.

\bibitem{Zen08}
Anil Zenginoglu.
\newblock {Hyperboloidal evolution with the Einstein equations}.
\newblock {\em Class. Quant. Grav.}, 25:195025, 2008.

\bibitem{VanHusHil14}
Alex Va{\~n}{\'o}-Vi{\~n}uales, Sascha Husa, and David Hilditch.
\newblock {Spherical symmetry as a test case for unconstrained hyperboloidal
  evolution}.
\newblock {\em Class. Quant. Grav.}, 32(17):175010, 2015.

\bibitem{Van15}
Alex Va{\~n}{\'o}-Vi{\~n}uales.
\newblock {\em {Free evolution of the hyperboloidal initial value problem in
  spherical symmetry}}.
\newblock PhD thesis, U. Iles Balears, Palma, 2015.

\bibitem{VanHus17}
Alex Vañó-Viñuales and Sascha Husa.
\newblock {Spherical symmetry as a test case for unconstrained hyperboloidal
  evolution II: gauge conditions}.
\newblock {\em Class. Quant. Grav.}, 35(4):045014, 2018.

\bibitem{HilHarBug16}
David Hilditch, Enno Harms, Marcus Bugner, Hannes R{\"u}ter, and Bernd
  Br{\"u}gmann.
\newblock {The evolution of hyperboloidal data with the dual foliation
  formalism: Mathematical analysis and wave equation tests}.
\newblock {\em Class. Quant. Grav.}, 35(5):055003, 2018.

\bibitem{GasHil18}
Edgar Gasper\'in and David Hilditch.
\newblock {The Weak Null Condition in Free-evolution Schemes for Numerical
  Relativity: Dual Foliation GHG with Constraint Damping}.
\newblock {\em Class. Quant. Grav.}, 36(19):195016, 2019.

\bibitem{DuaFenGasHil22}
Miguel Duarte, Justin~C Feng, Edgar Gasper\'in, and David Hilditch.
\newblock Regularizing dual-frame generalized harmonic gauge at null infinity.
\newblock {\em Classical and Quantum Gravity}, 40(2):025011, dec 2022.

\bibitem{Fri81}
H.~{Friedrich}.
\newblock {On the Regular and the Asymptotic Characteristic Initial Value
  Problem for Einstein's Vacuum Field Equations}.
\newblock {\em Proc. R. Soc. Lond. A}, 375(1761):169--184, March 1981.

\bibitem{Val16}
Juan-Antonio Valiente-Kroon.
\newblock {\em Conformal Methods in General Relativity}.
\newblock Cambridge University Press, Cambridge, 2016.

\bibitem{Fra04}
J{\"o}rg Frauendiener.
\newblock Conformal infinity.
\newblock {\em Living Rev. Relativity}, 7(1), 2004.

\bibitem{Fri95}
H.~Friedrich.
\newblock {Einstein equations and conformal structure - Existence of anti de
  Sitter type space-times}.
\newblock {\em J. Geom. Phys.}, 17:125--184, 1995.

\bibitem{LueVal12}
Christian Lübbe and Juan Antonio~Valiente Kroon.
\newblock The extended conformal einstein field equations with matter: The
  einstein{\textendash}maxwell field.
\newblock {\em Journal of Geometry and Physics}, 62(6):1548--1570, jun 2012.

\bibitem{Fri02}
Helmut Friedrich.
\newblock Spin-2 fields on minkowski space near spacelike and null infinity.
\newblock {\em Classical and Quantum Gravity}, 20(1):101--117, dec 2002.

\bibitem{LueVal09}
C.~L\"ubbe and J.~A. {Valiente Kroon}.
\newblock On de {S}itter-like and {M}inkowski-like spacetimes.
\newblock 26:145012, 2009.

\bibitem{GasVal17}
Edgar Gasperin and Juan~A. Valiente~Kroon.
\newblock Perturbations of the asymptotic region of the schwarzschild--de
  sitter spacetime.
\newblock {\em Annales Henri Poincar{\'e}}, pages 1--73, 2017.

\bibitem{DouFra16}
Georgios Doulis and J\"org Frauendiener.
\newblock Global simulations of minkowski spacetime including spacelike
  infinity.
\newblock {\em Phys. Rev. D}, 95:024035, Jan 2017.

\bibitem{Hub99}
Peter H{\"u}bner.
\newblock A scheme to numerically evolve data for the conformal {E}instein
  equation.
\newblock {\em Class. Quantum Grav.}, 16:2823--2843, 1999.

\bibitem{Hub01}
Peter H{\"u}bner.
\newblock From now to timelike infinity on a finite grid.
\newblock {\em Class. Quantum Grav.}, 18:1871--1884, 2001.

\bibitem{Hus02b}
Sascha Husa.
\newblock Problems and successes in the numerical approach to the conformal
  field equations.
\newblock volume 604 of {\em Lecture Notes in Physics}, pages 239--260.
  Springer, 2002.

\bibitem{Val03a}
J.~A. Valiente~Kroon.
\newblock Polyhomogeneous expansions close to null and spatial infinity.
\newblock In J.~Frauendiener and H.~Friedrich, editors, {\em The Conformal
  Structure of Spacetimes: Geometry, Numerics, Analysis}, Lecture Notes in
  Physics, page 135. Springer, 2002.

\bibitem{BeyDouFra13}
Florian Beyer, George Doulis, Jörg Frauendiener, and Ben Whale.
\newblock The spin-2 equation on minkowski background.
\newblock In {\em Springer Proceedings in Mathematics and Statistics}, pages
  465--468. Springer Berlin Heidelberg, sep 2013.

\bibitem{Pae13}
T.-T. Paetz.
\newblock Conformally covariant systems of wave equations and their equivalence
  to {E}instein's field equations.
\newblock {\em Ann. Henri Poincar\'{e}}, 16:2059, 2013.

\bibitem{CarHurVal}
Diego~A. Carranza, Adem~E. Hursit, and Juan A.~Valiente Kroon.
\newblock Conformal wave equations for the einstein-tracefree matter system.
\newblock {\em General Relativity and Gravitation}, 51(7), jul 2019.

\bibitem{Fri81a}
H.~Friedrich.
\newblock On the regular and the asymptotic characteristic initial value
  problem for {Einstein}'s vacuum field equations.
\newblock {\em Proc. R. Soc.}, 375:169, 1981.

\bibitem{Fri81b}
H.~Friedrich.
\newblock The asymptotic characteristic initial value problem for {Einstein}'s
  vacuum field equations as an initial value problem for a first-order
  quasilinear symmetric hyperbolic system.
\newblock {\em Proc. R. Soc.}, 378:401, 1981.

\bibitem{Faraoni1998qx}
Valerio Faraoni, Edgard Gunzig, and Pasquale Nardone.
\newblock {Conformal transformations in classical gravitational theories and in
  cosmology}.
\newblock {\em Fund. Cosmic Phys.}, 20:121, 1999.

\bibitem{Capozziello2011et}
Salvatore Capozziello and Mariafelicia De~Laurentis.
\newblock {Extended Theories of Gravity}.
\newblock {\em Phys. Rept.}, 509:167--321, 2011.

\bibitem{Clifton2011jh}
Timothy Clifton, Pedro~G. Ferreira, Antonio Padilla, and Constantinos Skordis.
\newblock {Modified Gravity and Cosmology}.
\newblock {\em Phys. Rept.}, 513:1--189, 2012.

\bibitem{Cho08}
Yvonne Choquet-Bruhat.
\newblock {\em {General Relativity and the Einstein Equations}}.
\newblock Oxford University Press, 12 2008.

\bibitem{Fri96}
H.~Friedrich.
\newblock Hyperbolic reductions for {E}instein's equations.
\newblock {\em Class. Quantum Gravit.}, 13:1451--1469, 1996.

\bibitem{Fri85}
H.~Friedrich.
\newblock On the hyperbolicity of {E}instein's and other gauge field equations.
\newblock {\em Comm. Math. Phys.}, 100:525--543, 1985.

\bibitem{LinSchKid05}
Lee Lindblom, Mark~A. Scheel, Lawrence~E. Kidder, Robert Owen, and Oliver
  Rinne.
\newblock A new generalized harmonic evolution system.
\newblock {\em Class. Quant. Grav.}, 23:S447--S462, 2006.

\bibitem{Wal84}
Robert~M. Wald.
\newblock {\em General relativity}.
\newblock The University of Chicago Press, Chicago, 1984.

\bibitem{Rainho_master_thesis}
Inês Rainho.
\newblock Linearized general relativity on hyperboloidal slices.
\newblock {\em Master Thesis.
  https://blackholes.tecnico.ulisboa.pt/gritting/theses.html\#section-5}, 2022.

\bibitem{PenRin84}
R.~Penrose and W.~Rindler.
\newblock {\em Spinors and space-time. {V}olume 1. {T}wo-spinor calculus and
  relativistic fields}.
\newblock Cambridge University Press, 1984.

\bibitem{PenRin86}
R.~Penrose and W.~Rindler.
\newblock {\em Spinors and space-time. {V}olume 2. {S}pinor and twistor methods
  in space-time geometry}.
\newblock Cambridge University Press, 1986.

\bibitem{GasVal20}
Edgar Gasperin and Juan Antonio~Valiente Valiente~Kroon.
\newblock {Zero rest-mass fields and the Newman-Penrose constants on flat
  space}.
\newblock {\em J. Math. Phys.}, 61(12):122503, 2020.

\bibitem{GasVal21a}
E~Gasper{\'{\i}}n and J~A~Valiente Kroon.
\newblock Staticity and regularity for zero rest-mass fields near spatial
  infinity on flat spacetime.
\newblock {\em Classical and Quantum Gravity}, 39(1):015014, dec 2021.

\bibitem{MinMacVal22}
Marica Minucci, Rodrigo~Panosso Macedo, and Juan~Antonio Valiente~Kroon.
\newblock The maxwell-scalar field system near spatial infinity, 2022.

\bibitem{ValAli22}
Mariem Magdy~Ali Mohamed and Juan A.~Valiente Kroon.
\newblock Asymptotic charges for spin-1 and spin-2 fields at the critical sets
  of null infinity.
\newblock {\em Journal of Mathematical Physics}, 63(5):052502, may 2022.

\bibitem{DuaFenGasHil23}
Miguel Duarte, Justin Feng, Edgar Gasperín, and David Hilditch.
\newblock The good-bad-ugly system near spatial infinity on flat spacetime.
\newblock {\em Classical and Quantum Gravity}, 40(5):055002, feb 2023.

\bibitem{Ste91}
J.~Stewart.
\newblock {\em Advanced general relativity}.
\newblock Cambridge University Press, 1991.

\bibitem{Hil15}
David Hilditch.
\newblock {Dual Foliation Formulations of General Relativity}.
\newblock arXiv:1509.02071, 2015.

\bibitem{MinVal23}
Marica Minucci and Juan Antonio~Valiente Kroon.
\newblock On the non-linear stability of the cosmological region of the
  schwarzschild-de sitter spacetime, 2023.

\bibitem{BarutElectrodynamics}
A.O. Barut.
\newblock {\em {Electrodynamics and Classical Theory of Fields \& Particles}}.
\newblock Dover Books on Physics Series. Dover Publications, 1980.

\bibitem{Mag07a}
Michele Maggiore.
\newblock {\em Gravitational Waves. Vol. 1: Theory and Experiments}.
\newblock Oxford University Press, Oxford, 2007.

\bibitem{Wheeler1955geons}
John~Archibald Wheeler.
\newblock Geons.
\newblock {\em Phys. Rev.}, 97:511--536, Jan 1955.

\bibitem{BrillHartle1964geon}
Dieter~R. Brill and James~B. Hartle.
\newblock Method of the self-consistent field in general relativity and its
  application to the gravitational geon.
\newblock {\em Phys. Rev.}, 135:B271--B278, Jul 1964.

\bibitem{AndersonBrill1997geon}
Paul~R. Anderson and Dieter~R. Brill.
\newblock Gravitational geons revisited.
\newblock {\em Phys. Rev. D}, 56:4824--4833, Oct 1997.

\bibitem{Sorkin:1981wd}
Rafael~D. Sorkin, Robert~M. Wald, and Zhen~Jiu Zhang.
\newblock {Entropy of self-gravitating radiation}.
\newblock {\em Gen. Rel. Grav.}, 13:1127--1146, 1981.

\bibitem{Schmidt:1999tr}
Heinz-Jurgen Schmidt and Felix Homann.
\newblock {Photon stars}.
\newblock {\em Gen. Rel. Grav.}, 32:919--931, 2000.

\bibitem{mitra2010likely}
Abhas Mitra and Norman~K Glendenning.
\newblock Likely formation of general relativistic radiation pressure supported
  stars or ‘eternally collapsing objects’.
\newblock {\em Monthly Notices of the Royal Astronomical Society: Letters},
  404(1):L50--L54, 2010.

\bibitem{Kim:2016jfh}
Hyeong-Chan Kim.
\newblock {Classifying self-gravitating radiations}.
\newblock {\em Phys. Rev. D}, 95(4):044021, 2017.

\bibitem{Hol22}
Jan~W. van Holten.
\newblock Curvature dynamics in general relativity, 2022.

\bibitem{Bie14}
Lydia Bieri and David Garfinkle.
\newblock Perturbative and gauge invariant treatment of gravitational wave
  memory.
\newblock {\em Physical Review D}, 89(8), apr 2014.

\bibitem{Cha83a}
S.~Chandrasekhar.
\newblock {\em The Mathematical Theory of Black Holes}.
\newblock Oxford University Press, Oxford, England, 1983.

\bibitem{xAct_web}
Jos{\'e}~M. Mart{\'i}n-Garc{\'i}a.
\newblock x{A}ct: tensor computer algebra.
\newblock \url{http://www.xact.es/}.

\end{thebibliography}


\end{document}